\documentclass{emulateapj}
\usepackage{changes}

\slugcomment{{\sc Accepted to ApJ} 2016 March 27}
\usepackage{graphicx}
\usepackage{longtable}

\def\lesssim{\mathrel{\spose{\lower 3pt\hbox{$\mathchar"218$}}
     \raise 2.0pt\hbox{$\mathchar"13C$}}}
\def\simlt{\mathrel{\spose{\lower 3pt\hbox{$\mathchar"218$}}
     \raise 2.0pt\hbox{$\mathchar"13C$}}}
\def\simgt{\mathrel{\spose{\lower 3pt\hbox{$\mathchar"218$}}
     \raise 2.0pt\hbox{$\mathchar"13E$}}}
\def\ltsima{$\; \buildrel < \over \sim \;$}
\def\gtsima{$\; \buildrel > \over \sim \;$}

\def\plottwo#1#2{\centering \leavevmode
\epsfxsize=.45\columnwidth \epsfbox{#1} \hfil
\epsfxsize=.45\columnwidth \epsfbox{#2}}

\shorttitle{Spectroscopy of a $z=3.78$ Protocluster}
\shortauthors{Dey et al. }
\begin{document}
\def\hh{\, h^{-1}}
\def\pcf1{PC~217.96+32.3}
\def\wrc4{{\it WRC4}}
\def\Msun{\rm M_\odot}
\def\kms{${\rm km\,s^{-1}}$}
\def\lya{${\rm Ly\alpha}$}

\title{Spectroscopic Confirmation of a Protocluster at $z\approx3.786$}
\author{
Arjun Dey\altaffilmark{1,2}, 
Kyoung-Soo Lee\altaffilmark{3}, 
Naveen Reddy\altaffilmark{4,5}, 
Michael Cooper\altaffilmark{6},
Hanae Inami\altaffilmark{1}, 
Sungryong Hong\altaffilmark{1,7}, 
Anthony H. Gonzalez\altaffilmark{8},
Buell T. Jannuzi\altaffilmark{9}}
\altaffiltext{1}{National Optical Astronomy Observatory, Tucson, AZ 85726}
\altaffiltext{2}{Fellow, Radcliffe Institute for Advanced Study, Harvard University, 10 Garden Street, Cambridge, MA 02138}
\altaffiltext{3}{Department of Physics and Astronomy, Purdue University, 525 Northwestern Avenue, West Lafayette, IN 47907}
\altaffiltext{4}{Department of Physics and Astronomy, University of California, Riverside, 900 University Avenue, Riverside, CA 92521}
\altaffiltext{5}{Alfred P. Sloan Fellow}
\altaffiltext{6}{Department of Physics and Astronomy, University of California, Irvine, CA}
\altaffiltext{7}{Department of Astronomy, University of Texas, Austin, TX}
\altaffiltext{8}{Department of Astronomy, University of Florida, Gainesville, FL 32611}
\altaffiltext{9}{Steward Observatory, University of Arizona, Tucson, AZ 85721}

\begin{abstract}
We present new observations of the field containing the $z=3.786$ protocluster, \pcf1. We confirm that it is one of the largest and most overdense high-redshift structures known. Such structures are rare even in the largest cosmological simulations. We used the Mayall/MOSAIC1.1 imaging camera to image a $1.2^\circ\times0.6^\circ$ area ($\approx150\times75$~comoving~Mpc) surrounding the protocluster's core and discovered 165 candidate \lya\ emitting galaxies (LAEs) and 788 candidate Lyman Break galaxies (LBGs). 
There are at least 2 overdense regions traced by the LAEs, the largest of which shows an areal overdensity in its core (i.e., within a radius of 2.5~comoving~Mpc) of $14\pm7$ relative to the average LAE spatial density ($\bar{\rho}$) in the imaged field. Further, $\bar{\rho}$ is twice that derived by other field LAE surveys. 
Spectroscopy with Keck/DEIMOS yielded redshifts for 164 galaxies (79~LAEs and 85~LBGs); 65 lie at a redshift of $3.785\pm0.010$. 
The velocity dispersion of galaxies near the core is $\sigma=350\pm40$~\kms, a value robust to selection effects. 

The overdensities are likely to collapse into systems with present-day masses of $>10^{15}\Msun$\ and $>6\times10^{14}\Msun$.  The low velocity dispersion may suggest a dynamically young protocluster. We find a weak trend between narrow-band (\lya) luminosity and environmental density: the \lya\ luminosity is enhanced on average by 1.35$\times$ within the protocluster core.  There is no evidence that the \lya\ equivalent width depends on environment. These suggest that star-formation and/or AGN activity is enhanced in the higher density regions of the structure.  \pcf1\ is a Coma cluster analog, witnessed in the process of formation.

\end{abstract}
  

\section{Introduction}

The formation and evolution of galaxies is known to be affected by the local environment in which they reside. At low redshift, galaxies in clusters have much lower star-formation rates than their field counterparts. Furthermore, various studies suggest that the present-day central ellipticals in clusters may have formed the bulk of their stars at high redshift and subsequently evolve passively \citep[e.g.,][]{stanford1998,vD2007,eisenhardt2008,mancone2010,snyder2012}. While this general picture is accepted, the detailed star-formation and assembly histories and quenching times of typical cluster galaxies and how these differ from those of field galaxies is not well understood (e.g., Snyder et al. 2012, Lotz et al. 2013). This is partly because systematic and robust searches for distant, young clusters are challenging at redshifts beyond $z\sim 2$ (e.g., Zeimann et al. 2012), due to both observational limitations and the decreasing number of physical systems that exist at these early epochs. 

One of the best ways to investigate galaxy formation in different environments is to witness it directly. To this end, several recent discoveries of very high-redshift (i.e., $z>3$) ``protocluster" regions are noteworthy, since these provide the ability of directly observing young galaxies forming in dense environments \citep[e.g.,][]{steidel2005,ouchi2005,venemans2007,overzier2008,diener2015,toshikawa2012,kuiper2012,lemaux2014,toshikawa2014,diener2015,casey2015,chiang2015}.  Although many of these discoveries rely on serendipity or the presence of a signpost (e.g., a distant quasar or radio galaxy), protocluster regions uniquely enable the study of galaxy properties in a range of environments. Kinematic studies of the member galaxies can be used to trace the dynamics of these growing overdense regions.

In \citet{lee2014}, we reported the discovery of an unusually dense region at a redshift $z\approx3.78$ containing three candidate protoclusters within a projected distance of $\approx$50~Mpc from each other. We designate this system \pcf1, based on the location of the most significant overdensity. The \pcf1\ structure was discovered serendipitously during a redshift survey, carried out using the DEIMOS spectrograph \citep{deimos} at the W. M. Keck Observatory, for UV-luminous Lyman Break Galaxies (LBGs) at redshift $z\sim 3.8$ \citep[see][for details]{lee2013} in the Bo\"otes field of the NOAO Deep Wide-Field Survey \citep[NDWFS;][]{NDWFS}. The initial spectroscopy resulted in the discovery of 5 LBGs with redshifts $z=3.783\pm0.002$ lying within 1~Mpc (physical distance) of each other \citep{lee2014}. In 2012 we obtained imaging through a narrow-band filter (with a bandpass which samples Ly$\alpha$ emission at $z=3.792\pm0.017$) of a $36^\prime\times36^\prime$ field using the MOSAIC1.1 camera at the Mayall 4m telescope on Kitt Peak, resulting in the discovery of 65 
Lyman Alpha Emitting galaxy (LAE) candidates within a comoving volume of $72\times72\times25$~Mpc \citep{lee2014}. The peak overdensity was found to be at the southern-most edge of the narrow-band survey field. 

In this paper we report on new spectroscopic observations of the LAE candidates discovered by \citet{lee2014} and on new narrow- and broad-band imaging data (extending to the south of the original narrow-band field imaged by Lee et al.~2014) that reveal the larger extent of the protocluster \pcf1\ and allow us to better quantify the overdensity. We describe our observations in \S2 and present the main analyses and results in \S3. We discuss the resulting implications for the protocluster region in \S4 and summarize our main findings in \S5. 

Throughout, we use  the WMAP7 cosmology $(\Omega, \Omega_\Lambda, \sigma_8, h)=(0.27, 0.73, 0.8, 0.7)$ from \citet{wmap7}; at $z=3.785$, the angular scale is 7.343~kpc arcsec$^{-1}$ (physical) or $\approx$126.5~Mpc deg$^{-1}$ (comoving). Distance scales are presented in units of comoving Mpc unless noted otherwise. All magnitudes are given in the AB system \citep{oke1983}. 

\begin{figure}[t]
\epsscale{1.2}
\plotone{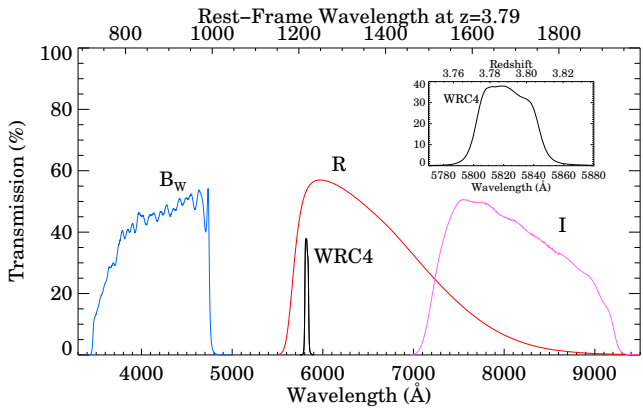}
\caption{The total throughput of the filters used to image the \pcf1\ field and identify Ly$\alpha$ emitters. The throughput shown includes the filter bandpass (see {\tt{https://www.noao.edu/kpno/mosaic/filters/filters.html}}) and have been multiplied by the response of the camera, optics (corrector and telescope) and atmospheric extinction. Atmospheric absorption due to O$_2$ and H$_2$O is not shown. The inset shows a detail of the \wrc4\ narrow-band filter transmission (in a converging f/3 beam) along with the redshift range of galaxies over which it samples the Ly$\alpha$ emission line. 
\label{filters}}
\end{figure}

\section{Observations \& Reductions}

\subsection{Imaging}

\begin{figure}[t]
\epsscale{1.2}
\plotone{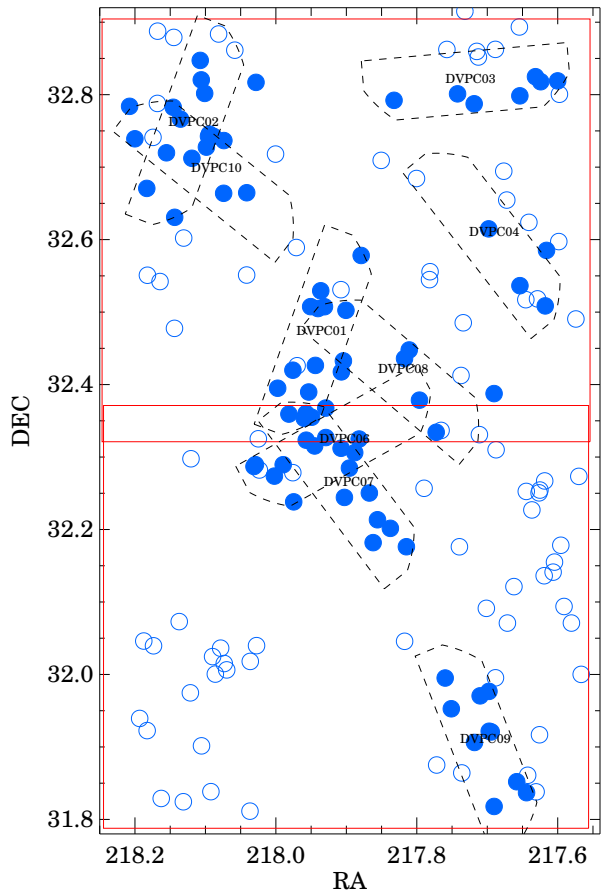}
\caption{The approximate location of imaging fields (red lines) and spectroscopic masks (dashed lines) overlaid on the spatial distribution of LAE candidates (circles) in the the field of the z=3.78 protocluster \pcf1. The region shown covers $\approx 150 \times 75~h^{-1}\,{\rm Mpc^2}$ (comoving). The DEIMOS masks are labelled. Filled circles show the spectroscopically confirmed LAEs. 
\label{imspecfields}}
\end{figure}

We used the MOSAIC1.1 camera \citep[]{mosaic1p1} at the f/3.1 prime focus of the 4m Mayall Telescope at the Kitt Peak National Observatory on U.T.~2014 April 24-29  to obtain deep narrow- and broad-band imaging of a field just south of the region imaged by \citet{lee2014}. The new field (central pointing of RA=14:31:36.14 DEC=+32:08:46.29, J2000) overlaps the previously imaged area by 9~arcmin, the overlap region including the most significant overdensity. Figure~\ref{imspecfields} shows the location of the two imaged fields (red squares). The field was imaged through the \wrc4\ narrow-band filter (hereafter \wrc4; KPNO filter \# k1024, $\lambda_{\rm cen}=5819$\,\AA, $\Delta\lambda=42$\,\AA), and the $B_W$, $R$ and $I$ broad-band filters used by the NDWFS \citep{NDWFS}. The filter transmission curves are shown in Figure~\ref{filters}. We used individual exposure times of 20~min, 20~min, 10~min, and 10~min in the \wrc4, $B_W$, $R$ and $I$ bands respectively, and the telescope was offset in random directions by $\approx$1-3~arcmin between exposures. The conditions were variable through the run. The seeing ranged between 0.9\arcsec\ and 2\arcsec\ with a median value of 1.4; two nights were clear and the rest had varying amounts of high cirrus. 
The total exposure times are summarized in Table~\ref{imobs}.

\begin{figure*}[ht]
\epsscale{1.2}
\plotone{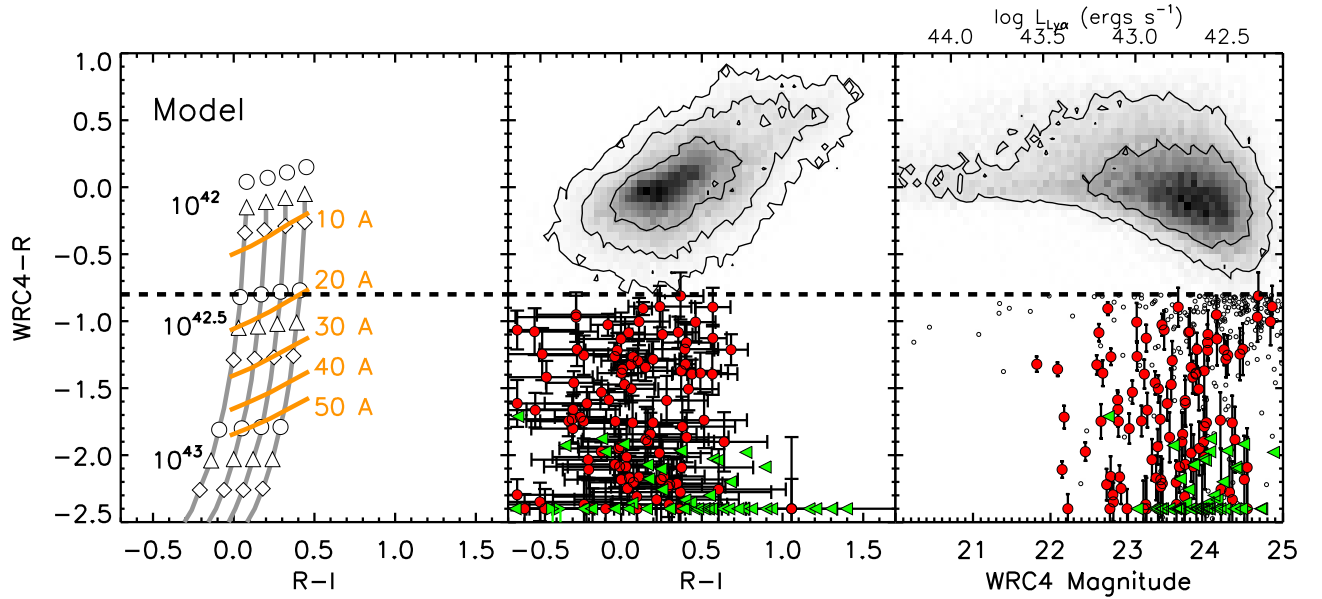}
\caption{
{\it Left:}  Predicted tracks for LAEs of different luminosities and reddening in the $WRC4-R$ vs $R-I$ color-color diagram \citep[see][for details]{lee2014}. The grey lines show the expected color evolution with Ly$\alpha$ luminosity (which increases from top to bottom and labelled on the left of the tracks) at four different reddening values. Circles, triangles, and diamonds represent three different $R$-band continuum magnitudes of 25.2, 25.5, and 25.8 respectively. Orange lines represent the Ly$\alpha$ rest-frame equivalent widths of (from top to bottom) 10, 20, 30, 40, and 50\AA\ at different continuum $R-I$ colors. The $WRC4-R$ color cut is chosen to match (approximately) the expected color of $W_0=20$\AA\ sources at the median $R-I$ color.  
{\it Middle:} The color-color data for all $WRC4$-band detected sources in the $1^\circ \times 0.5^\circ$ field surrounding \pcf1. Photometric LAE candidates selected according to equation 1 (i.e., below the thick dashed line) are shown as circles ($S/N(R)\geq 2$) or the green, left-facing triangles ($S/N(R)<2$, i.e., undetected in $R$). Sources that are formally undetected in the $I$-band are represented as upper limits. Galaxies with $(WRC4-R)\leq -2.4$ are shown at the color position of $-2.4$. The upper limits for the sources with ${\rm S/N} (R, I) < 2$ (i.e., undetected in $R$ or $I$) are shown as triangles. 
{\it Right:} The color-magnitude diagram for the detected sources. Sources that fail the $(B_W-R)$ color cut are shown as small open circles. The approximate Ly$\alpha$ luminosities corresponding to the \wrc4\ magnitude are indicated on the upper abscissa; these values are calculated for LAEs with an $R$-band magnitude fixed to the median observed value of  25.5. 
\label{lae_selection}}
\end{figure*}

The data were reduced using the NOAO pipeline up to the stage of bias- and overscan-corrected and dome- and sky-flattened frames with associated bad pixel masks. Distortion corrections and astrometry for the MOSAIC frames were computed using objects from the SDSS DR7 catalog \citep{sdssdr7}. The individual images were then projected to a tangent plane matching that of the reprocessed NOAO Deep Wide-Field Survey Bo\"otes field broad-band data, and image stacks were constructed. We also reprocessed the NDWFS data in the same manner: we remeasured the image distortions and astrometric zero points using the SDSS catalog and reprojected and stacked the data to create new $B_W$, $R$ and $I$-band frames. 
The photometric calibration of the $B_W$, $R$ and $I$ band data was done by tying to the NDWFS photometry; $WRC4$ band data was photometrically calibrated using observations of spectrophotometric standard stars \citep{specphot1,specphot2}. 
We direct the reader to \citet{lee2014} for further details about the reduction and calibration of the MOSAIC imaging data. 

The final image stacks in the southern field have 5$\sigma$ depths (in a 2\arcsec\ diameter aperture) of 26.96, 26.34, 25.57 and 24.91~AB mag in the $B_W$, $R$, $I$ and \wrc4 bands respectively. These depths are comparable to those attained in the northern field \citep[26.88, 26.19, 25.37, 24.86~AB mag;][]{lee2014}. The stacked images were spatially registered and object catalogs for the stacked images were constructed using Source Extractor \citep{sextractor}. 

\subsection{Spectroscopy}

We obtained spectroscopy of the LAE candidates
at the W. M. Keck II Telescope on the nights of U.T.~2014 May 6 and 2015 April 18 (see Table~\ref{specobs}). We used the DEIMOS spectrograph \citep{deimos} equipped with the 600ZD grating (600~lines~mm$^{-1},\ \lambda_{\rm blaze}=7500$\AA) and slitlets of width 1.1~arcsec, corresponding to a spectral resolution FWHM of $\approx$6\AA. The grating was tilted to cover the wavelength range 4900\AA $ - 1\mu$m (approximately) and used in conjunction with a BAL12 filter. 
On U.T.~2014 May 6, there was some high cirrus and the seeing was $\approx$0.6\arcsec\ for the first 2/3 of the night  and variable between 0.8 and 1\arcsec\ for the last part of the night.  On U.T.~2015 April 18, the observations were obtained under clear skies; the seeing was $\approx0.7\arcsec-1.0\arcsec$\ during most of the night, but deteriorated to $\approx1\arcsec-1.3\arcsec$ for the last two hours.  Over the two nights, we observed 9 masks targeting 100 LAE candidates, 184 $B_W$-drop candidates and various filler sources (see Table~\ref{specobs}). The approximate locations and orientations of the DEIMOS masks are shown in figure~\ref{imspecfields}.

Wavelength calibration was performed using a combination of NeArKrXe and HgCdZn arc lamps, and checked using the telluric emission lines. 
The RMS scatter in the line centroids of the [\ion{O}{1}]$\lambda$5577 telluric line is $<$0.13\AA, implying that errors in the wavelength calibration result in a scatter of $<10^{-4}$ in redshift.
The relative spectrophotometric calibration was performed using observations of the spectrophotometric standards Feige~34 and Wolf~1346 \citep{specphot1,specphot2}.
The spectroscopic data were reduced using the DEIMOS DEEP2 pipeline \citep[``spec2d'';][]{spec2d1,spec2d2}. During the analyses of the 2014 May 6 data, we discovered that the data had an unexplained offset in the wavelength calibration, which we traced back to an offset in the Flexure Compensation System that occurred during our run. M. Cooper modified the pipeline in order to correct for this offset and derive the correct wavelength calibrations. The final wavelength calibrations were verified using the telluric lines.  

Spectroscopy of a few galaxies associated with \pcf1\ was obtained using the Hectospec spectrograph on the MMT telescope on U.T.~2012 June 1,2. These observations were taken as part of a large spectroscopic survey targeting bright $B_W$ dropout candidates with the goal of finding very luminous ($>3L^*$) Lyman Break Galaxies at $z\sim 3.8$. Hectospec was equipped with 270 l/mm grating ($\lambda_b=5200$\AA), which provides a resolution of $\approx$1.2\AA/pixel. The Hectospec observations yielded the redshift of 1 additional LAE at the redshift of the protocluster lying near the core of the NW group. 

\section{Results \& Analysis}
\subsection{Photometric Selection and Spectroscopic Confirmation of Ly$\alpha$ Emitting Galaxy Candidates}

\begin{figure*}[hbt]
\epsscale{0.85}
\vspace{-0.3in}
\plotone{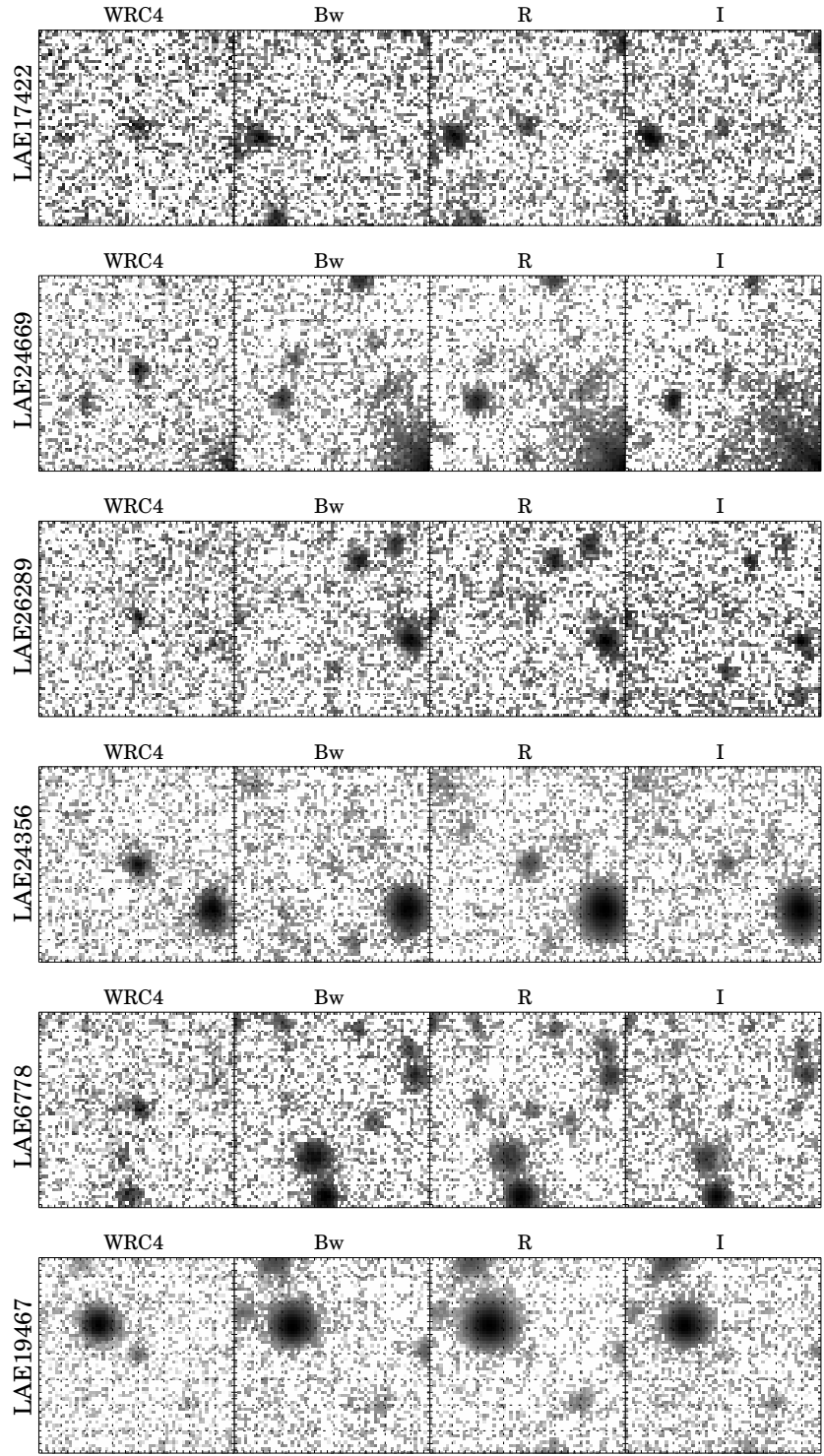}
\caption{Sample cutout images of LAE candidates in the \wrc4, $B_W$, $R$ and $I$ bands. Each panel is 15\arcsec\ ($\approx$110~kpc) on a side, centered on the LAE candidate. The greyscale stretch is logarithmic. \label{images}}
\end{figure*}

We selected candidate LAEs at $3.769<z<3.804$ (i.e., the approximate filter bandpass) using the same criteria described by \citet{lee2014}:
\begin{eqnarray}
(WRC4-R) <-0.8 ~~ \cap  ~~~S/N(WRC4)\geq 10 ~~~~~~~~~ \nonumber \\
\cap ~~ \{(B_W-R)>1.8 ~~\ \ \cup~~ S/N(B_W) < 2\}. ~~~~~~~~~~~~~~
\end{eqnarray}
Figure~\ref{lae_selection} shows the color-color and color-magnitude distributions of the detected sources and the main selection criteria. The left panel shows the relationship between the \wrc4\ magnitude and ($WRC4-R$) color as a function of the Ly$\alpha$ luminosity ($L_{\rm Ly\alpha}$), rest-frame equivalent width ($W_{\rm Ly\alpha}$), and reddening (please see Lee et al.~2014 for details regarding the spectral models used for this simulation). 
The selection described resulted in 165 LAE candidates distributed over the $\approx 150\times75~h^{-1}\,{\rm Mpc^2}$ (comoving) field. The photometry of these candidates is presented in Table~\ref{laephot}. Sample cutout images of the LAE candidates are shown in Figure~\ref{images}.

\begin{figure*}[hbt]
\epsscale{1.0}
\plotone{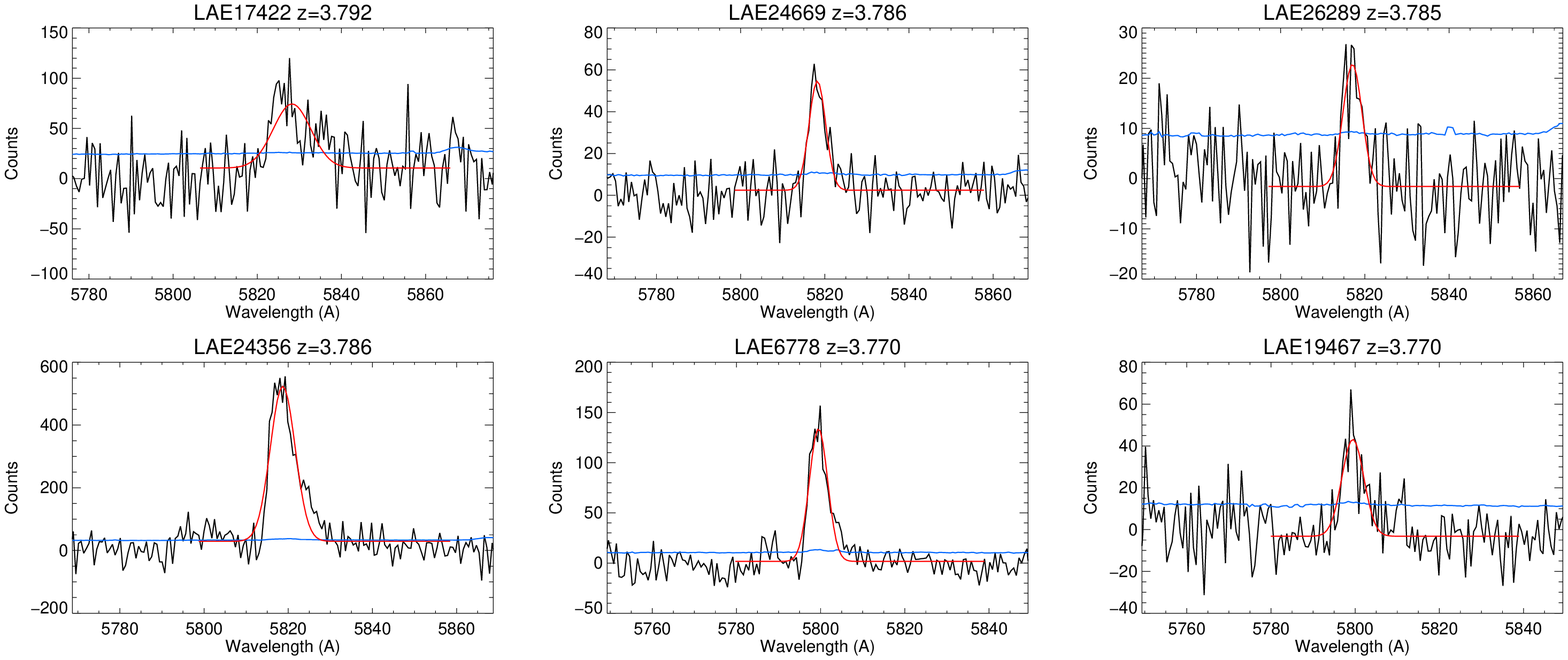}
\caption{Sample DEIMOS spectra of the LAEs showing the detection of the Ly$\alpha$ emission line. The red line shows a model gaussian fit to the line. The blue line shows the level of the 1$\sigma$ noise spectrum. \label{spectra}}
\end{figure*}

Of the 100 LAE candidates observed spectroscopically through 9 DEIMOS masks, we detected emission lines in 81 of them; 79 were found to lie at $3.76<z<3.81$ and 2 are most likely [\ion{O}{2}] emitters at low redshift. Of the 19 sources that did not yield a line detection: four failures are due to problems with the slitlets (e.g., because of targets being masked by the slitmask support structure or errors in the mask design); two are due to irrecoverable problems with sky subtraction; and only 13 of the sources are true failures, where no emission line was detected despite the candidate being observed. Of these 13, 10 occurred on the three masks with $\le$1 hour of exposure time, all of which were taken at high airmass, suggesting that the failures are largely due to under-exposure of the masks. The remaining 3 candidates may be the result of poor photometry. The overall success rate of the LAE selection is 79/94 or $\approx81\%$; counting just the well-exposed masks (i.e., $t_{\rm exp}\ge5400$s), the success rate is 62/70, or $\approx$89\%. 
None of the spectra of the confirmed LAEs show evidence of strong AGN activity (i.e., no broad lines or strong lines of high ionization species).
Sample spectra of six of the LAE candidates are presented in Figure~\ref{spectra}. 

The redshift yield for the $B_W$-dropout Lyman break galaxies (LBGs) was much poorer than that of the LAEs: of the 184 targeted $B_W$-dropout LBG candidates, 71 yielded robust redshifts and an additional 8 yielded tentative redshifts.  The yield is also lower than that achieved by our original $B_W$-dropout selection presented in \citet{lee2013}; the difference is because the new masks targeted significantly fainter candidates and with shorter total exposure times per mask.  Of the 71 robust redshifts, 2 are low-redshift interlopers and the remainder span a broad redshift distribution $3.4<z<4.2$ \citep{lee2011}; 23 of the confirmed LBGs lie within the range [3.75,3.85].  Forty of the spectroscopically confirmed LBGs show Ly$\alpha$ in emission. 

\begin{figure}[hbt]
\epsscale{1.3}
\hspace{-0.8in}
\plotone{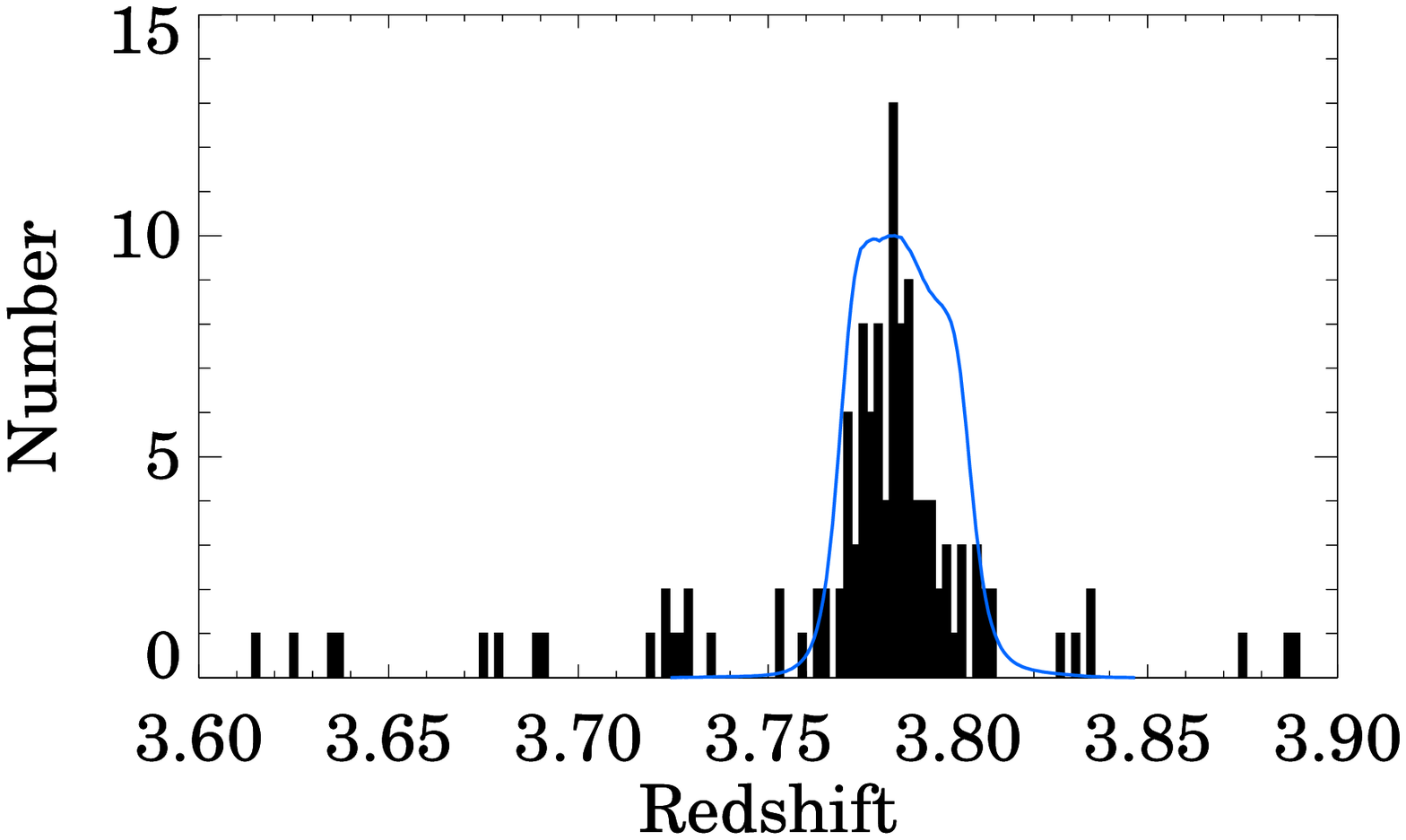}
\caption{The histogram shows the redshift distribution based on 81 LAE and 44 LBG redshifts in the redshift range $3.6\le z\le 3.9$ in the $1.2^\circ\times0.6^\circ$ field surrounding \pcf1,  
The blue line shows, for comparison, the transmission bandpass (with arbitrary normalization) for the \wrc4\ narrow-band filter used to select the LAEs. \label{zhist}}
\end{figure}

For the LAE candidates we measured redshifts from gaussian fits to the Ly$\alpha$ emission line. The redshifts of the LAEs are presented in Table~\ref{laespec}. Given the slit losses due to seeing, galaxy size, and guiding errors, we decided to estimate the Ly$\alpha$ flux and luminosity from the narrow-band imaging data, correcting for the continuum flux contamination in the narrow-band photometry using measurements (or 3$\sigma$ lower limits) of the observed equivalent width from the spectra. Since the narrow-band filter covers only $\approx$42\AA, the line flux $F_{\rm Ly\alpha}$ is related to the total flux in the narrow-band filter $F_{WRC4}$ as follows: 
\begin{equation}
F_{\rm Ly\alpha} \approx F_{WRC4}\left(1+{{\Delta\lambda}\over{W_{\rm Ly\alpha}^{\rm obs}}}\right)^{-1}
\end{equation}
where $W_{\rm Ly\alpha}^{\rm obs}$ is the observed equivalent width of the Ly$\alpha$ emission line and $\Delta\lambda$ is the width of the filter. The measured rest-frame equivalent widths and implied line fluxes and luminosities are also listed in Table~\ref{laespec}. 
The final redshift histogram within the region is shown in Figure~\ref{zhist}.

\subsection{Characterization of the Large Scale Structure}

The deeper imaging observations show that the structure discovered by \citet{lee2014} extends further to the south. The main overdensity (labelled as the ``S group'' in \citet{lee2014} and now located in the center of our 0.5$^\circ\times1^\circ$ field) is more fully traced, better quantified and found to be more overdense than previously thought. There is marginal evidence for the overall structure being extended from the NE to SW corners of the imaged field, suggestive of a $>$170~Mpc sized filament. The field also shows regions which are devoid of LAE candidates, the largest of these located near [$\alpha,\delta$]=[217.92,+31.9], and measures $\approx$30~Mpc in diameter. 

\subsubsection{Overdensity and Mass Estimates\label{sec:overdensity}}

The new imaging data not only elucidate the larger extent of the structure, but also provide a better measure of the average field density. The 165 LAE candidates are distributed over $\approx$2236~arcmin$^2$. Since the total area includes the protocluster regions, we estimated the average surface density 
by excluding the regions around the two main overdensities (i.e., the central and NE regions): we find 127 LAE candidates within an area of 2097~arcmin$^2$, corresponding to an average field density of  
$(6.06\pm0.54)\times 10^{-2}$~LAE per arcmin$^2$ ($0.0136\pm0.0012~{\rm Mpc^{-2}}$).  
For comparison, we expect 64$\pm$8 LAE in the same area, based on integrating the Ly$\alpha$ luminosity function derived by \citet{ouchi2008} and imposing our narrow-band selection criteria \citep[see][for details]{lee2014}. Hence the overall field surrounding \pcf1\ (excluding the two most overdense regions), $\approx\times75~{\rm Mpc}^2$ (comoving) in size, is overdense by a factor of nearly 2.

\begin{figure}[hbt]
\epsscale{1.25}
\plotone{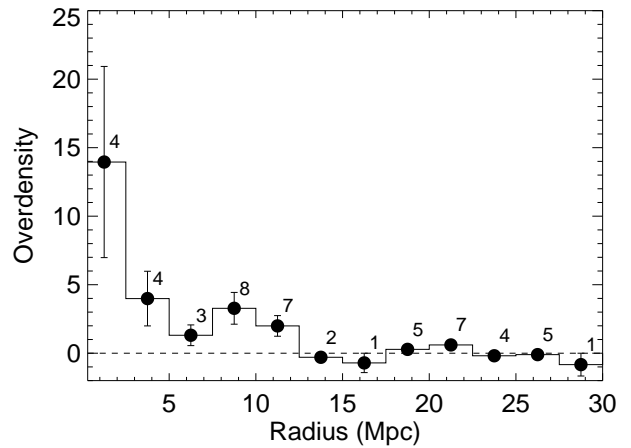}
\caption{The overdensity of LAEs, computed in concentric circular annuli of width 2.5~Mpc, as a function of distance from the center of the protocluster (at [$\alpha,\delta$]=[217.965$^\circ$,32.35$^\circ$]). The mean density of the field is computed from the LAEs lying within 20~Mpc and 60~Mpc of the center. Uncertainties shown are derived assuming Poisson statistics. The central region ($<2.5$~Mpc) is overdense by a factor of $14\pm7$. 
A wider region within a radius of 12~Mpc is overdense by a factor of $\approx4.5$. 
Each bin is labelled with the number of LAE candidates within the annulus. 
\label{overdensity}}
\end{figure}

Figure~\ref{overdensity} shows the variation in density (adopting our measured field density) measured in circular annuli centered on the core of \pcf1\ at ($\alpha,\delta$)=(217.965$^\circ$,32.35$^\circ$). The cluster core within a radius of 2.5~Mpc (comoving) contains 4 LAEs and corresponds to an areal overdensity ($\equiv (\Sigma(<r)/\Sigma_{\rm field}-1)$) of $\approx14\pm7$. Nineteen galaxies lie within a radius of 10~Mpc from the core, corresponding to an average areal overdensity of $3.5\pm1.0$.  
If the estimate based on the \citet{ouchi2008} luminosity function is a better measure of the field density, then the overdensity values derived here should be approximately doubled (i.e., $29\pm15$ and $7.8\pm2$, respectively). The \citet{ouchi2008} study used a broader filter and detected 101 LAEs in a single 0.95~deg$^2$ field, only 26 of which were spectroscopically confirmed.  Given the differences in the observational set-up, we conservatively estimate overdensities based on the field density measured self-consistently from our data. Given that our field density estimate is derived in an overdense field, our resulting overdensity estimates should be considered strong lower limits. 

In cosmological simulations, (proto-)cluster sizes are estimated as the mass-weighted average distance of all subhalos that will end up in a fully virialized descendant. Using this definition, Chiang et al. (2013) estimated that the typical extent of protoclusters at $z\sim 4$ is $R_e\approx9.0\pm 1.4$ comoving Mpc for Coma cluster analogs, and $\approx 6.2\pm1.3$ comoving Mpc for Virgo cluster analogs. Our observations have not sampled all the cluster members, nor do we know the typical halo mass of the LAEs; hence, we can only make an analogous estimate of the size. The area within which the surface density is 3$\times$ the average density for the central core is $\approx$78 arcmin$^2$ (345 Mpc$^2$) and the (projected) effective radius is $\approx$10.5 Mpc. This is qualitatively comparable in size to the Chiang et al. prediction for Coma analogs. However, a robust quantitative comparison to the models is not possible given the current data.

To estimate the total enclosed (present-day) masses of the protocluster systems, we use two approaches: one utilizing the correlation between galaxy overdensity and total mass measured by \citet{chiang2013} from the Millennium I and II Simulation Runs \citep{springel2005}; and the other based on a more heuristic approach of considering the magnitude of the overdensity and the bias of the LAE population. The details of our implementation of both methods are presented in \citet{lee2014}.

First, we measure galaxy overdensity in the Core and NE clump regions within a cylindrical volume defined by a circular area 13~Mpc (6.17~arcmin) in diameter and height corresponding to the depth of the narrow-band survey ($\approx$20~Mpc), centered on the positions [$\alpha$,$\delta$]=[217.944, 32.395] and [218.144,32.748], respectively. The size scale is chosen to match that of the Chiang et al.~study, in which the overdensity of protoclusters was measured from the Millennium simulation \citep{springel2005} in a cubic tophat of volume $15\times15\times15$~Mpc$^3$. We estimate the galaxy overdensity as $\delta_g=(N_{\rm{group}}/N_{\rm{field}})-1$ where $N_{\rm{field}}$ is the mean number of field LAEs within a 13~Mpc diameter circle. As described above, we conservatively estimate our field LAE density to be $(6.06\pm0.54)\times 10^{-2}$~arcmin$^{-2}$, and thus $N_{\rm{field}}$=1.81. 
Within the same circular area, we find 12 LAEs and 11 LAEs in the Core and NE regions, respectively, corresponding to galaxy overdensities of $\delta_g=5.6^{+0.6}_{-0.5}$  and $5.1^{+0.6}_{-0.5}$, respectively. Interpolating from Fig.~10 of \citet[]{chiang2013}, the measured overdensities imply $(0.6-1.3)\times 10^{15}~M_\odot$ and $(3-8)\times10^{14}~M_\odot$ for the Core and NE group, respectively. The scatter measured for the masses of simulated protoclusters in their sample is roughly 0.2~dex. 
If, instead, we adopt the estimate of the field density derived from the \citet{ouchi2008} luminosity function, we calculate overdensities of 12 and 11 in the Core and NE regions respectively. In the \citet[]{chiang2013} study, there is no cluster with a galaxy overdensity exceeding $\delta_g = 9$. 

An alternative estimate of the mass uses the approach described by \citet{steidel1998} and is based on estimating the total mass within a galaxy overdensity that will virialize by the present epoch \citep[see also][]{lee2014}. This estimate relies on a measure of the scale and volume of an overdensity and the inherent bias of the galaxy population. We consider the volumes enclosed by the relative overdensity $\delta_g$=3 isodensity contours in Figure~7. The $\delta_g$=3 contour is chosen since it appears to contain the bulk of the galaxies within each of these regions (29 and 14 in the Core and NE regions, respectively). The areas enclosed by these contours surrounding the Core and NE clump regions are approximated well by circular areas of radii 13~Mpc and 10~Mpc, respectively. Assuming that the galaxy bias for the LAE population is $b_{\rm LAE}=2.0$ \citep{gawiser2007,lee2014}, we estimate the enclosed masses as $1.0\times 10^{15}M_\odot$ and $6.2\times 10^{14}M_\odot$ for the Core and NE group, respectively. These numbers are in agreement with the previous mass estimates based on the comparison to the Millennium simulations.

\begin{figure}[ht]
\epsscale{1.2}
\plotone{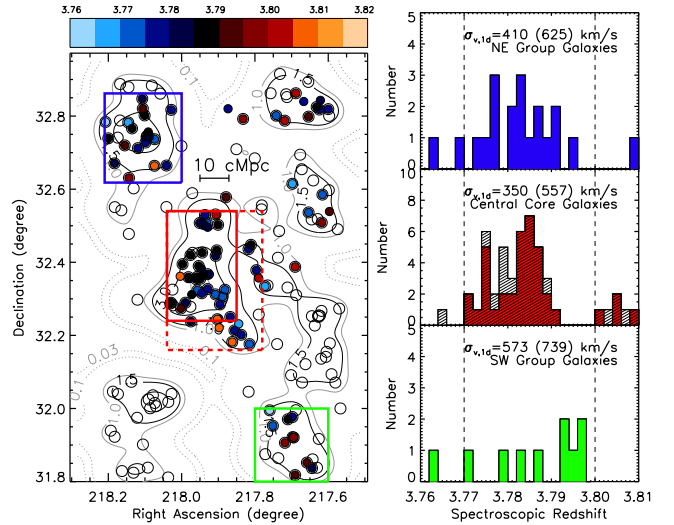}
\caption{{\bf Left:} The spatial distribution of LAE candidates (circles) in the the field of the z=3.78 protocluster \pcf1. The region shown covers $\approx 150 \times 75~h^{-1}\,{\rm Mpc^2}$ (comoving). The contours are constructed by smoothing the positions of the candidate LAEs and spectroscopically confirmed member LBGs with a gaussian kernel of FWHM$\approx$10~Mpc. LAEs with spectroscopically measured redshifts are represented as solid points, with colors representing their redshift as shown in the color bar. The colored rectangles enclose LAEs in the Core (red),  NE group (blue), and SW group (green).  {\bf Right:} Redshift distributions of the LAEs in the three rectangular regions: NE group (upper panel), Core (middle panel) and SW group (lower panel). The vertical dashed lines are separated by $\Delta z$=0.03, which corresponds to $\approx$25~Mpc front-to-back. In the central panel, the red histogram corresponds to the galaxies within the solid red rectangle in the left panel, and the grey histogram adds the galaxies between the solid and dashed red rectangles. The histograms are labelled with the velocity dispersions computed both for the galaxies within the redshift region $3.77<z<3.80$ demarcated by the vertical dashed lines and (in parentheses) for the galaxies in the larger redshift range of $3.76<z<3.81$. 
The median velocity in each region appears to show a monotonic large scale gradient from the NE to SW and may reflect the overall velocity gradient within a filamentary structure.
\label{zclump}}
\end{figure}

The field density estimates are affected by Poisson variance in the galaxy counts, and cosmic variance. We estimate the net uncertainty by randomly sampling the LAEs within the field, and find that the total fractional error, $\delta N/N_{\rm{field}}$, is 0.9.  This translates into a similar fractional uncertainty in the galaxy overdensity. 

There are two main uncertainties in the mass estimate described above: (i) the volume containing the mass overdensity; and (ii) the bias factor which relates the measured galaxy overdensity to underlying matter overdensity. As for the former, adopting a lower overdensity for mass estimate would lead to a larger volume, resulting in a roughly similar enclosed mass. If we adopt $\delta_g = 2$, the enclosed transverse area is 731~Mpc$^2$ and 450~Mpc$^2$, for the Core and NE regions, respectively, corresponding to the enclosed masses of $1.14\times 10^{15}M_\odot$ and $7.0\times 10^{14}M_\odot$. Thus, the uncertainties arising from a specific choice of overdensity contours is 10\%. As for the latter, galaxy bias is a function of halo mass and environments. Based on the measurement of LAE clustering, Gawiser et al. (2007) estimated the LAE bias to be $1.7^{+0.4}_{-0.3}$ at $z\sim3$. We independently carried out cell variance of our LAEs as a measure of LAE bias, and determined $b=2.0\pm0.2$. While our estimate is consistent with that of Gawiser et al. (2007), it is slightly lower than the the value derived from the Millennium Runs, $b\sim 2.35$, for galaxies with star formation rate $>1M_\odot~\rm{yr}^{-1}$. Using the bias $b=2.35$ (1.7) would result in the masses $9.6\times10^{14}M_\odot$ ($1.1\times 10^{15}M_\odot$) and $5.7\times10^{14}M_\odot$ ($6.7\times10^{14}M_\odot$), for the Core and NE group, respectively, 8\% lower (10\% higher) than our earlier estimate using the bias of $2$. 

The mass estimates are very large given the area imaged by our narrow-band imaging study. We repeated the analyses of the Millennium Simulation presented by \citet{lee2014} using our increased search volume of $150\times72\times20~{\rm Mpc}^3$, and find that the median number of progenitors of clusters of mass $>10^{14}\Msun$ is one. The likelihood of finding {\em two} structures, the sum of whose masses is $>1.5\times10^{15}\Msun$ is 3\%; the likelihood of a single structure with mass $>10^{15}\Msun$ is 1\%. 

Both mass estimates presented here are conservative underestimates, since we are measuring the overdensities relative to the local field density. As noted, this is twice the average field density of LAEs measured by other field surveys \citep[e.g.,][]{ouchi2008}. The mass estimates suggest that this protocluster region will collapse into a cluster similar to mass of Coma in the present epoch.

\begin{figure}[t]
\epsscale{1.2}
\plotone{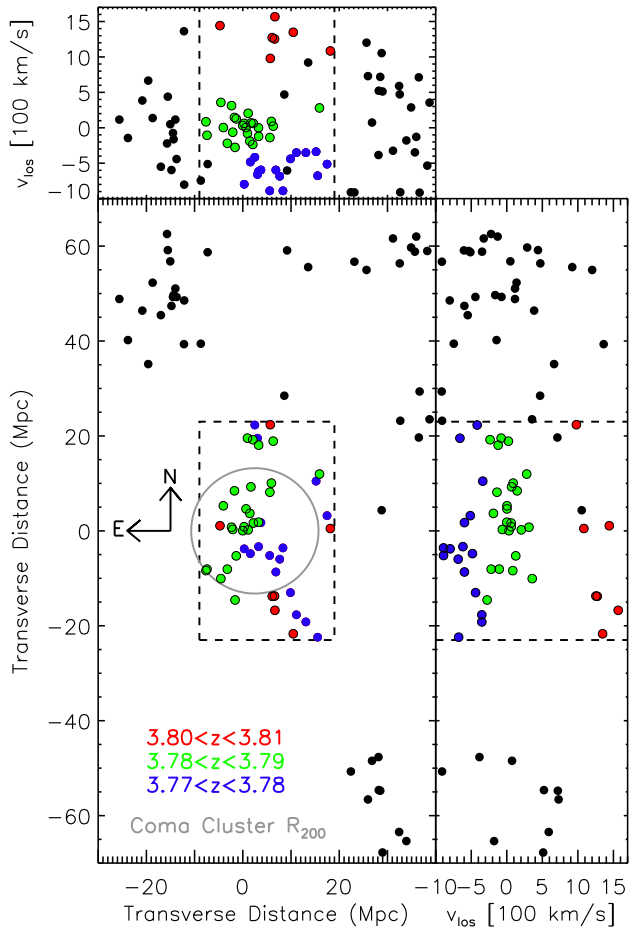}
\caption{
{\bf Left:} The three-dimensional spatial distribution of spectroscopically confirmed LAEs in the $z=3.78$ protocluster region. Each dot shows an LAE with a measured redshift. The top and side panels show projections in velocity space relative to the adopted central redshift of the structure of 3.785 (i.e., $\Delta v = (z-3.785)c/4.785$~\kms). The red, green and blue dots show three narrow redshift ranges within the rectangular region covering the central protocluster core. The grey circle denotes the virial radius ($R_{200}$) for the Coma cluster \citep[1.99$h^{-1}$~Mpc scaled to the appropriate comoving angular size scale;][]{kubo2007}, shown for comparison. The central region appears to be made up of at least two groups of LAEs (denoted by the blue and green dots) that are separated in redshift by $\approx$500~\kms. 
\label{3dskyplot} }
\end{figure}

\subsubsection{Velocity Structure}

The spectroscopic observations presented here confirm that the photometrically identified LAEs lie within a narrow redshift range $3.76<z<3.81$ as expected (see Figure~\ref{zhist}). The data also show that two of the overdensities identified by \citet{lee2014} -- the central and the NE regions -- show small velocity spreads (see Figure~\ref{zclump}), suggesting that these LAEs mark physical structures and not line-of-sight projections. The similarity in the median redshift in both regions suggests that they are part of a coherent large scale structure that extends over many tens of Mpc.  The galaxy distribution in the central region is elongated along the north-south direction, with most galaxies lying in an oblong core, and a few lying in a semi-detached clump to the north.  

We subdivide the structure into four spatial rectangular regions as follows: 

\begin{centering}
\begin{tabular}{ll}
Core:       & 217.78 $\le\alpha\le$ 218.04 ; 32.16 $\le\delta\le$ 32.54 \\
Inner Core: & 217.85 $\le\alpha\le$ 218.04 ; 32.24 $\le\delta\le$ 32.54 \\
NE Clump:   & 218.00 $\le\alpha\le$ 218.20 ; 32.62 $\le\delta\le$ 32.86 \\
SW Region:  & 217.60 $\le\alpha\le$ 217.80 ; 31.80 $\le\delta\le$ 32.00 \\
\end{tabular}
\end{centering}

\noindent The regions are shown in Figure~\ref{zclump}; the Inner Core region is a subset of the Core region. Spectroscopic observations have yielded 48 redshifts in the core region, shown by the red rectangle in Figure~\ref{zclump}. Thirty-nine of these confirmed systems lie within the narrow redshift range 3.774--3.790 (middle right panel of Figure~\ref{zclump}). The NE overdensity, centered at roughly ($\alpha$,$\delta$)=(218.15,+32.75), contains 18 members with spectroscopically measured redshifts, all but two lie within the redshift range 3.775--3.796. 

In each of these regions, we estimated the line-of-sight velocity dispersion from the measured spectroscopic redshifts using four different techniques: simple standard deviation from the mean; median absolute deviation from the sample median (M.A.D.); the ``gapper" scale estimator which is derived from considering the distribution of gaps in the ordered statistics; and the biweight scale estimator \citep[see][for details]{beers1990}. The gapper and biweight methods have been shown to be more robust against a few outliers and for non-gaussian underlying distributions, with the former method preferred for sample sizes $<10$. The resulting estimates (using both galaxies within the core regions and those within the wider overdensities) are presented in Table~\ref{veldisp}.  We estimated the uncertainties in these measures using bootstrap analysis, drawing samples with replacement from the data.

The line of sight velocity dispersions are small, given the magnitude of the overdensities (see \S~\ref{sec:overdensity} and Lee et al. 2014),  in both the central region ($\approx$350~\kms) and the NE region ($\approx$430~\kms). The SW region which shows the least coherence and has the fewest redshift measurements shows a larger dispersion of $\approx$580~\kms. Based on the larger dispersion, small number of measured redshifts, and lack of coherence, we cannot confirm that the SW group is a physical structure.

The use of rectangular regions is driven by simplicity. However, all the galaxies included in the redshift histograms shown in the right panels of figure~\ref{zclump} lie within well-defined overdensity contours. Choosing, for example, only galaxies within the $\delta_g=1.5$ overdentisy contour does not change the results. Given that the overdensity contours are themselves uncertain, being derived from the as-yet-spectroscopically unconfirmed photometric LAE sample, we resort to the simple definition of rectangular regions for our analyses in this paper.

The velocity dispersion estimates presented here are distinguished from other protocluster dispersion measurements in the literature by the large number of spectroscopically measured redshifts in relatively small areas, e.g., 34 redshifts within the Core region which projects an area of $\approx$34~Mpc$^2$ (physical) on the sky. The large number of redshifts allow us to investigate the velocity structure of the system in more detail. Figure~\ref{3dskyplot} shows the three-dimensional distribution of the galaxies with spectroscopic redshifts in a narrow velocity range ($\delta v \approx 2500$~\kms; $3.769\le z \le 3.809$) around \pcf1. Within the central region, the LAEs fall into three possible velocity groupings along the line of sight, with the two richest groups (members shown by the blue and green dots) separated by $\approx$500~\kms\ (see also Figure~\ref{zclump}). While the two groups do overlap along the line of sight, the lower redshift one (blue dots) lies a bit further to the southwest. These two groups may represent merging subgroups within the overdensity, providing further circumstantial evidence the structure is dynamically young.

The velocity dispersion measurements presented here are dominated by the LAEs, which have been photometrically selected in a fairly narrow redshift range ($\sim 2000$~\kms). However, Monte Carlo tests with small numbers of galaxies distributed uniformly in redshift throughout the range sampled by the narrow-band filter confirm that it is unlikely that the small velocity dispersion results from the limited bandpass. For example, for samples of 43 galaxies distributed randomly in redshift, velocity dispersions less than 400~\kms\ occur with $<$0.002\% probability. 
It is also unlikely that we are catching the velocity edge of the protocluster, since the LAE redshift peaks are centered in the bandpass and the LAE number density is already significantly in excess of what is expected based on the \citet{ouchi2008} luminosity function.
The use of LAEs with redshifts measured from Ly$\alpha$ can also result in an overestimate of the true velocity dispersion, since the Ly$\alpha$ emission line centroid can be biased by the kinematics of the gas in the galaxies (i.e., outflows and inflows). \citet{erb2014} studied the velocity offsets between Ly$\alpha$ and the systemic galaxy redshift (as measured from the rest-frame optical nebular emission lines) and found that the velocity offset correlates with continuum luminosity, with less luminous galaxies showing smaller offsets. Using their entire sample, we find that the dispersion in the velocity offset is $\sigma(\Delta v_{\rm Ly\alpha})\approx154$~\kms. Correcting for this effect would reduce the measured velocity dispersion of the \pcf1\ core region to $\approx$315~\kms.
In addition, the velocity dispersion estimates assume that all the LAEs along the line of sight are part of the structure. However, if any of the LAEs are foreground or background contaminants (i.e., lying at the near or far ends of the redshift range selected by the narrow-band filter) rather than associated with a self-gravitating structure, their contribution would bias the protocluster velocity dispersion to higher values. The measured velocity dispersion may therefore be an upper limit to the true dispersion of the structure.

\begin{figure}[t]
\epsscale{1.4}
\hspace{-0.7in}
\plotone{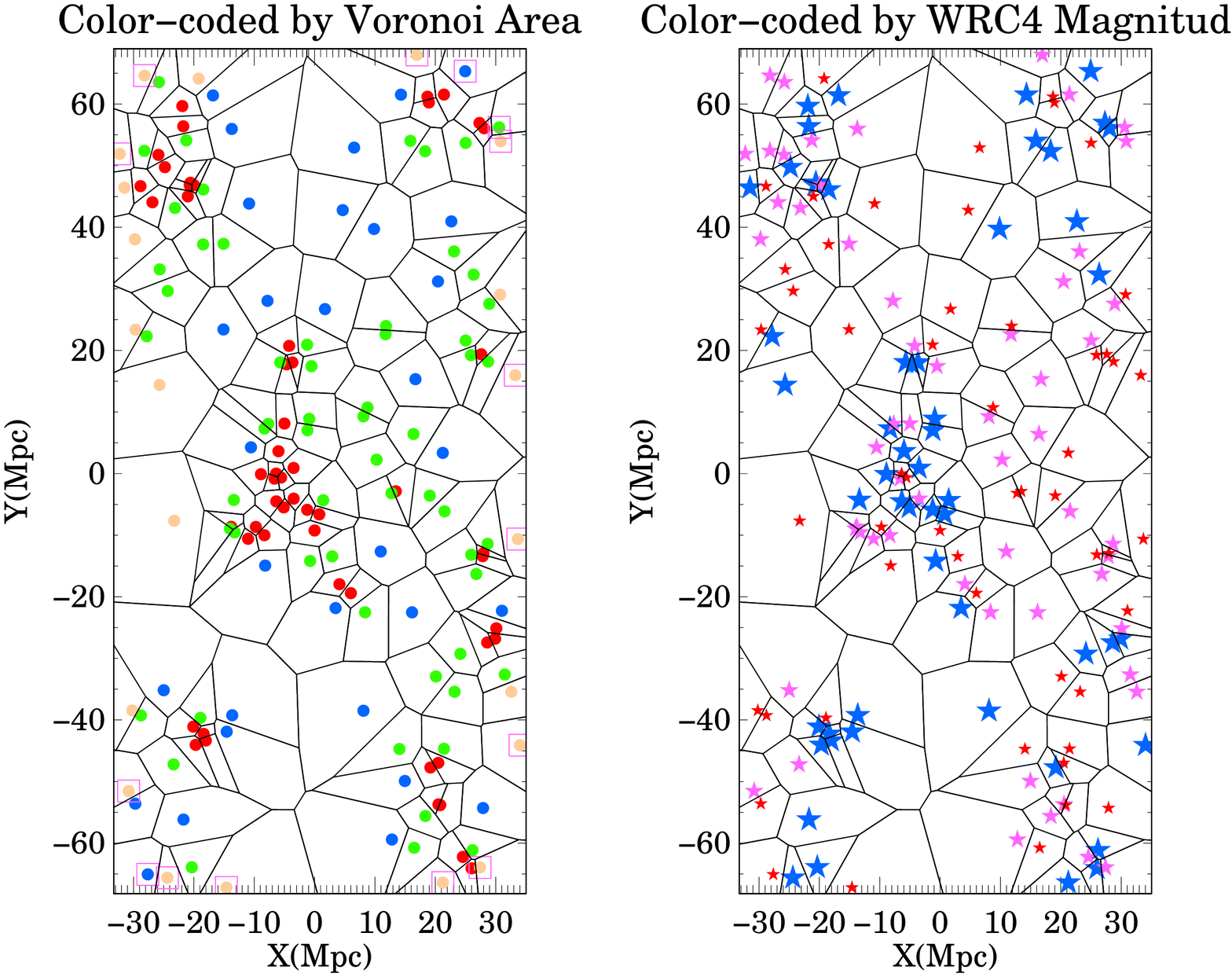}
\caption{Voronoi tesselation of the 2-d spatial distribution of all photometrically selected LAE candidates. In the left panel, the LAEs occupying equivalent circular areas with radii $r_{V}<3$, $3\le r_{V}<5$, $5\le r_{V}<10$ are shown as red, green and blue dots respectively. The LAEs on the boundary of the survey area (and hence with unbounded Voronoi polygons) are enclosed by purple squares. The area of each polygon is related to the inverse of the local density. In the right panel, the LAEs are divided by \wrc4\ magnitude, with the magnitude ranges $<$23.5, 23.5$-$24, and $>$24.0 denoted by large blue, green, and small purple stars. There is a tendency for brighter narrow-band emitters to lie within denser regions. \label{voronoi}}
\end{figure}

\subsection{Environmental Effects on Ly$\alpha$ Emission}

In order to study the dependence of the Ly$\alpha$ emission properties on the local environment, we estimated the local environmental density for each photometrically-selected LAE candidate as the inverse of the area of its associated Voronoi polygon \citep{marinoni2002,cooper2005}. We then characterize the (inverse) density by the radius of the equivalent circular area that is equal to the area of the Voronoi polygon, i.e., $r_{V}=\sqrt{(A_{V}/\pi)}$. The Voronoi tesselation of the LAE distribution is shown in Figure~\ref{voronoi}; the left panel shows the division of the LAEs into three density classes: $r_{V}<3$~Mpc; $3\le r_{V}<5$~Mpc; and $5\le r_{V}<10$~Mpc. Sources with $r_{V}\ge10$~Mpc are considered boundary sources with unbounded Voronoi polygons. 

As a proxy for the Ly$\alpha$ luminosity, we use the narrow-band magnitude measured through the \wrc4\ filter. The front-to-back depth of the LAE sample due to the bandpass of the filter only results in an $<$4\% uncertainty on the luminosity. A larger uncertainty in the luminosity may result from the contribution of continuum flux in the \wrc4\ bandpass, but for rest-frame Ly$\alpha$ equivalent widths $\ge$25\AA, the correction to the flux is $<$25\%. The mean $(WRC4-R)$ color for the sample is $-1.9$, suggesting that the correction for the continuum contamination is typically $\sim$15\%.

\begin{figure}[ht]
\epsscale{1.3}
\hspace{-0.5in}
\plotone{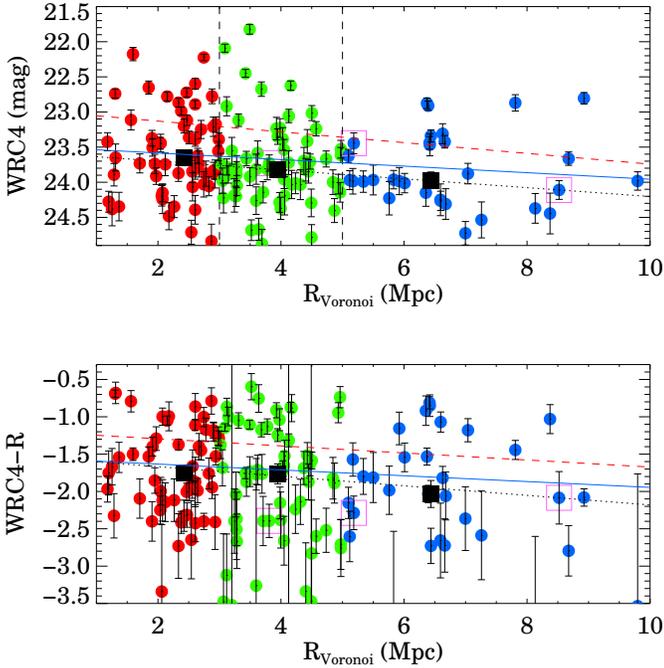}
\caption{The variation of \wrc4\ magnitude (related to the Ly$\alpha$ line luminosity; top panel) and $WRC4-R$ color (related to the Ly$\alpha$ line equivalent width; bottom panel) versus the radius of the circle enclosing the Voronoi area (related to the inverse sqare root of the local density). The colored dots represent the emitters in the three different $r_{V}$ ranges described in the caption and left panel of Figure~\ref{voronoi}. The solid black squares represent the median values in each $r_{V}$ range. Dots that are enclosed by purple squares are emitters near the boundary of the survey where the $r_{V}$ is not well determined because the Voronoi polygons are unbounded. The dotted black, dashed red and solid blue lines represent, respectively, least absolute deviation, linear least-squares, and a Bayesian approach to linear regression fits to the data.  There is a tendency for the most luminous emitters and the bluer emitters to lie in denser regions, although the scatter is large.\label{lyavsRv}}
\end{figure}

The right panel of Figure~\ref{voronoi} shows the spatial distribution of LAEs as a function of \wrc4\ magnitude. The variation of \wrc4\ magnitude (i.e., Ly$\alpha$ luminosity) and $WRC4-R$ color (roughly Ly$\alpha$ equivalent width) with $r_V$ are shown in Figure~\ref{lyavsRv}.  Despite significant scatter, there is a weak trend of brighter \wrc4\ magnitudes (i.e., more luminous Ly$\alpha$ emission, toward denser regions.  The most luminous emitters (with \wrc4\ mag $<$23.0) appear to be predominantly in regions with $r_{V}<4$~Mpc and trace the core structures in the field. Lower luminosity LAEs are visible in both overdense and underdense regions. In contrast, the ($WRC4-R$) color shows no significant variation with $r_V$. This suggests that while Ly$\alpha$ luminosity may correlate with density, the Ly$\alpha$ equivalent width does not. 

To quantify the significance of the trend of luminosity with $r_V$, we applied the Bayesian method of linear regression developed by \citet{kelly2007}. We find a slope of $\Delta m_{WRC4}/\Delta r_V \approx +0.05$~mag Mpc$^{-1}$, with a 4\% probability of the slope being $\le 0$. 
The 95\% (68\%) confidence limits on the slope are [$-0.01$, 0.11] ([0.02,0.08]). The slope for the variation of color versus $r_V$ is $\approx -0.04$~mag Mpc$^{-1}$ with 95\% / 68\% confidence ranges of [$-0.10$,0.02] / [$-0.07$,0.01] and a 11\% probability of a zero or positive slope. We conclude that the trend with $r_V$ of Ly$\alpha$ luminosity is marginally significant, whereas that of color is not (see Figure~\ref{lyavsRv}). 

\begin{figure}[t]
\epsscale{1.3}
\hspace{-0.3in}
\plotone{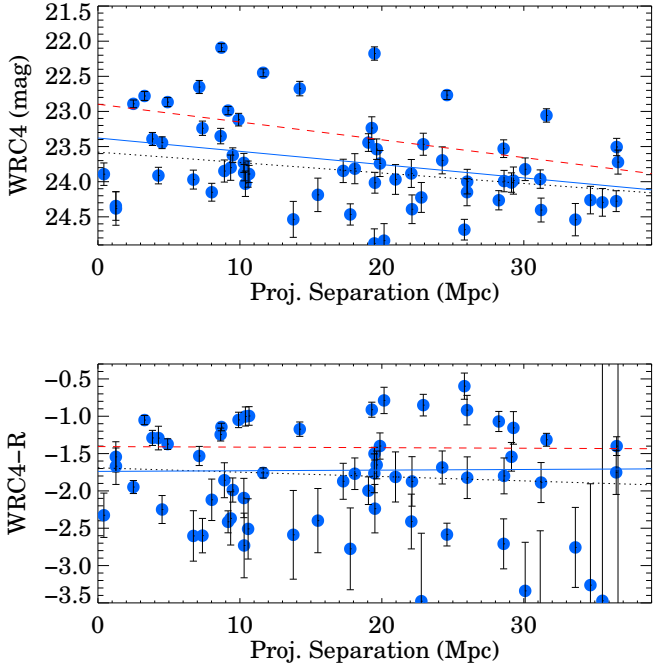}
\caption{The distribution of \wrc4\ narrow-band magnitudes (related to the Ly$\alpha$ line luminosity; top panel) and narrow-band color excess (related to the Ly$\alpha$ line equivalent width; bottom panel) for the LAE candidate sample vs the projected distance from the central protocluster core (i.e., the highest density region in Figure 1, located at $14.53^h,32.35^d$). 
As in Figure~\ref{lyavsRv}, the dotted black, dashed red and solid blue lines represent, respectively, least absolute deviation, linear least-squares, and Bayesian approach to linear regression fits to the data. 
LAEs at $r_P<12$~Mpc appear to have a higher median narrow-band flux (i.e., higher median Ly$\alpha$ luminosity) than the LAEs at larger projected distances. There is no dependence of color on $R_P$. A clearer trend may emerge with a larger sample of field and protocluster galaxies once membership has been established and line luminosity accurately measured. \label{envplot}}
\end{figure}

We also investigated whether the LAE properties vary with distance
from the center of the protocluster \pcf1 (defined to lie at
[$\alpha,\delta$]=[14.531,+32.350]). We find a weak dependence of
\wrc4\ magnitude with protocluster distance (Figure~\ref{envplot}).
Using the \citet{kelly2007} Bayesian approach, we find that the
trend of luminosity with $R_{\rm proj}$ has a preferred slope of
$\approx 0.087$~mag Mpc$^{-1}$, with a 95\% (68\%) confidence range
of [0.041,0.167] ([0.069,0.133]). In contrast, the ($WRC4 - R$) color shows
no significant variation with  $R_{\rm proj}$ (slope of 0.004~mag Mpc$^{-1}$
with 95\%/68\% confidence ranges of [$-$0.076,0.060]/[$-$0.042,0.028]).  These results are
consistent with the preceding analysis (using the Voronoi tesselation)
that more luminous Ly$\alpha$ emitters tend to lie in denser regions,
but with Ly$\alpha$ equivalent width showing no dependence on
environmental density.   In other words, both the Ly$\alpha$ line
emission and the UV continuum emission in LAEs appears to be enhanced
in denser regions.

%

\section{Discussion}


\subsection{The True Overdensity of PC~217.96+32.3}

The very large overdensity observed in the LAE distribution is not seen in the Millennium simulation. While this could imply that the protocluster region is unusually rare, it is possible that LAEs are more biased tracers of the large scale matter distribution. Indeed, we observe the Ly$\alpha$ emission to be enhanced in denser regions. The median value of the \wrc4\ magnitude in the denser regions (i.e., with $r_V\le 3.0$) is 0.32~mag brighter than that in the $3.0\le r_V<10$~Mpc region, suggesting that the Ly$\alpha$ luminosity is enhanced by a factor of $\approx1.35$ in protocluster regions. If the Ly$\alpha$ luminosity traces the star-formation rate, then this is a further indication that SFRs are enhanced in denser regions at high redshift. If this bias occurs uniformly across LAEs of all luminosity, the implication is that the LAE spatial distribution exaggerates overdense regions, and measures of the overdensity from (line-flux of continuum limited) LAE samples will be biased by this factor. Whether this luminosity bias also affects the presumably older and more luminous LBG samples remains to be seen. Correcting the Core region of the PC217.96+32.3 protocluster for this factor of 1.35 bias suggests that the true areal overdensity is $\approx10-11$ ($\approx3.3$) within a radius of 2.5~Mpc (10~Mpc). A larger statistical study sampling Ly$\alpha$-emitting galaxies both field and protocluster environs will be important to understand this bias and thereby estimate the true galaxy overdensity.

\subsection{Velocity Dispersion and Extent of  Protoclusters}

PC~217.96+32.3 is an usually massive structure consisting of multiple subgroups, each characterized by a surprisingly low velocity dispersion ($\approx 350$~\kms). While we have already established that such structures are extremely rare, it is also of interest to investigate how unusual PC~217.96+32.3 is compared to other known protoclusters. Here, we compare PC~217.96+32.3 with the best studied high-redshift protocluster systems, primarily focused on the spatial extent and velocity dispersion measurements. 

The total line of sight velocity dispersion measured using {\underline{all}} the spectroscopically confirmed LAEs in the field of view is 880$\pm$160~\kms. However, this number is almost certainly biased due to the nature of our selection (see Appendix). 
The line of sight velocity dispersions of two overdense regions in PC~217.96+32.3 (the central and NE regions) are small compared with those estimated for many other protocluster candidates. For example, in a large protocluster candidate identified by a 7$\times$ overdensity at $z=6.01$, \citet{toshikawa2014} spectroscopically confirmed  28 galaxies and measured a very high velocity dispersion of $\approx870\pm85$~\kms\ \citep[see also][]{toshikawa2012}.  The origin of these differences is not clear: it could be 
because (a) other protocluster systems are significantly more massive; (b) the spectroscopic redshift samples for many protocluster candidates include interloping galaxies that are not true members; (c) the PC~217.96+32.3 protocluster system is less evolved dynamically than most other candidates; or (d) the $z=3.78$ system is non-spherical and seen along the minor axis of its velocity ellipsoid. In an attempt to converge on a likely reason we consider the best studied protocluster systems. 

Many of the known protoclusters are found as a result of targeted searches for LAEs around high redshift radio galaxies.
\citet{venemans2007} surveyed the vicinities of eight radio galaxies and discovered six protoclusters at $2.0<z<5.2$, with sizes $>$1.75~Mpc (physical) and masses of $(2-9)\times 10^{14}~{\rm M_\odot}$
\citep[see also][]{venemans2005}. The authors measured velocity dispersions for these clusters ranging from $\approx250$~\kms\ to 1000~\kms, with the velocity dispersions being lowest for the two highest redshift clusters. 
In addition, the two protoclusters with the highest measured velocity dispersions, MRC~0052$-$241 and MRC~1138$-$262, show bimodal redshift distributions: MRC~0052$-$241 has peaks with velocity dispersions of 185~\kms\ and 230~\kms; and MRC~1138$-$262 has peaks with velocity dispersions of 280~\kms\ and 520~\kms. TN~J1338$-$1942 at $z=4.1$ is estimated by \citet{venemans2007} to be the most massive structure in their sample, with $M\approx (6-9)\times 10^{14}M_\odot$; yet, it has a relatively low velocity dispersion similar to that measured in the core of PC~217.96+32.3. 
One possible interpretation is that these protocluster regions are in the process of coalescence, similar the Core of PC~217.96+32.3 (Fig.~\ref{3dskyplot}). MRC~0052-241 may be in an earlier stage of merging as the two subgroups are separated by $\sim1500$~\kms.  

The protocluster SSA22a at $z=3.09$ is one of the largest, best-studied system at high redshift \citep{steidel1998}. Similar to  PC~217.96+32.3, a redshift spike of luminous LBGs led to its initial discovery and the followup study revealed a large number of LAEs in the same large-scale structure \citep{steidel2000,hayashino2004}. The spectroscopically confirmed LBG members span a 14~Mpc$\times$10~Mpc region, within which roughly $10^{15}~M_\odot$ of total mass is enclosed. The LAEs trace a belt-like large-scale structure centered on the LBGs, spanning $\approx 40$~Mpc in its longest dimension. Using 16 LBG redshifts published by Steidel et al. (1998; their Table~1), we compute a velocity dispersion of $866\pm115$~\kms. \citet{matsuda2005} reported the velocity dispersion of $\sim 1100$~\kms\ for their LBG/LAE sample. The spectroscopic LAEs appear to have at least three separate groups, one at $z=3.075$, $z=3.095$ and $z=3.105$ (1500~\kms\ and 730~\kms in velocity space) where the spectroscopic LBGs are shared between the two latter groups (see Figure~1 of Matsuda et al. 2005). Although we are unable to recompute the velocity dispersions of these subgroups (the redshifts are not published), based on the figures in \citet{matsuda2005} we estimate that they range over  400-500~\kms. Similar to PC~217.96+32.3, SSA22a appears to consist of multiple groups. The substructure in the spatial and velocity distributions may suggest that these systems are dynamically young and in the process of merging.

Recently, three different studies have obtained redshifts of galaxies within a single large protocluster region at $2.428<z<2.487$ (median $z\approx2.467$) located within the COSMOS field at 150.076+2.355 \citep[]{diener2015,chiang2015,casey2015}. These studies jointly identified 66 galaxies lying within a region $\approx 0.4^\circ$ ($\sim 40$ comoving Mpc) in size. The region in COSMOS has many similarities with the higher redshift protocluster region discussed here. The distribution of galaxies is spread over a wide area with varying density: there is a high density region \citep[covered by the][studies]{diener2015,chiang2015} and a more diffuse and wide spread filamentary region \citep[covered by the][study]{casey2015}.  
We have recomputed the velocity dispersions for the entire structure using all published redshifts and find that the wider region has a velocity dispersion of $\approx$1300~\kms, significantly higher than that of the protocluster region presented in this paper. Restricting the sample to the 20 galaxies lying within the region studied by \citet{diener2015} and \citet{chiang2015}, we measure velocity dispersions of $\approx$700~\kms. The higher values of the velocity dispersion may indicate that the COSMOS field structures are more dynamically evolved than \pcf1. It is also possible that the COSMOS and \pcf1\ structures are similar in size and dynamical state, but viewed from different perspectives, with the COSMOS structure distributed along our line of sight (and therefore yielding a higher line-of-sight velocity dispersion) and \pcf1\ distributed transverse to our line of sight.

Based on these considerations, we conclude, perhaps not surprisingly, that velocity dispersion measured from a handful of galaxies is not a reliable measure of the mass of a protocluster as it is only loosely related to the total mass of an unvirialized system. The velocity dispersion measurement may be compromised by the clumpiness of the protocluster region, its state of dynamical evolution, and, perhaps, the observer's perspective. In particular, we caution readers about deriving any conclusions about the evolution in velocity dispersion with redshift since velocity dispersion measurements can be biased for several reasons. First, lower velocity dispersion is expected at higher redshift if samples are selected at different redshifts using narrow-band filters of the same width (see Appendix). Second, for unvirialized systems such as protoclusters, measured velocity dispersion will depend on the dynamical maturity of a structure and  on the observers' viewing angle.

\subsection{Implications of the Environmental Dependencies}

A \wrc4\ trend with environmental density (or distance from the cluster core) suggests that the Ly$\alpha$ luminosity may be enhanced in denser regions. The lack of a trend of ($WRC4-R$) with distance from the cluster core suggests that the UV luminosity is also enhanced in denser regions, i.e., that the Ly$\alpha$ equivalent width is not strongly dependent on local density. If the Ly$\alpha$ and UV luminosity are both tracers of the star-formation rate, then the observations suggest that the star-formation is enhanced in denser regions at these redshifts in contrast to what is observed for present-day clusters. Indeed, observations of clusters at $z\sim 1-1.8$ have shown that the star-formation - environmental density relation may be in the process of reversing at $z>1.5$; the observations presented suggests that this process has fully reversed by $z\sim3-4$ \citep[e.g.,][]{elbaz2007,cooper2008,hilton2009,tran2010,brodwin2013,zeimann2013,alberts2014,wagner2015,santos2015}. 

The LAEs may therefore be simply exhibiting the trend observed for the more luminous galaxy population in intermediate redshift galaxy clusters. If the luminosity is due to star-formation, then the observations imply that the rate of star-formation is enhanced within dense environments at high redshift, perhaps because the accretion rate efficiencies are higher or that the merger rate between gas-rich dwarfs is enhanced. 
If the Ly$\alpha$ emission is indeed enhanced in dense environments, then finding overdensities of LAEs may be a good way to identify high-redshift protoclusters. The Ly$\alpha$ emitting phase in these galaxies may be short-lived \citep[e.g.,][]{partridge1967} or their visibility may be modulated by the covering fraction of gas \citep[e.g.,]{trainor2015}, but LAEs also tend to be drawn from the more abundant population at the low-mass end of the luminosity function.

At slightly lower redshifts ($z\sim 2-3$), the evidence for environmental dependencies is mixed with different studies resulting in varying results even within the same field. For example, the COSMOS field has a protocluster system at $z\approx$2.45 spread over a $\approx0.4^\circ$ region that has been targeted by multiple groups \citep[]{diener2015,casey2015,chiang2015}. In one region within this field \citet{diener2015} report a possible increase in the number of more massive and quiescent galaxies relative to the field fractions. In contrast, \citet{casey2015} find an enhanced number of rapidly star-forming galaxies and AGN associated with a neighboring (i.e., within $\sim50-100$~Mpc) protocluster core at $z$=2.47, lying $\sim50-100$~Mpc away from the \citet{diener2015} region. \cite{chiang2015} identify another cluster of LAEs at $z=2.44$ in a region in between the \citet{diener2015} and \citet{casey2015} cores and report a higher median stellar mass and a possible excess of AGN within this third region. 

Since no homogeneous data on the Ly$\alpha$ fluxes or star-formation rates are publicly available for the spectroscopically confirmed galaxies in COSMOS $z\approx2.44$ protocluster region, we cannot assess whether the field exhibits similar trends in $L_{\rm Ly\alpha}$ with environmental density. However, it is clear that the COSMOS protocluster region should be investigated for any possible trends, as it is similar in extent and overdensity to the protocluster region reported in this paper. 

At higher redshifts, no strong evidence for environmental dependencies has been reported. However, this may be due in large part to the limited samples thus far investigated, the lack of sufficient spectroscopy to exclude interlopers, and the lack of any constraints (resulting from an error analysis) on what trends can be excluded \citep[e.g.,][]{cooper2007,cooper2010}. In the $z\approx6.01$ protocluster candidate region studied by \citet{toshikawa2012,toshikawa2014}, the authors find no evidence for any environmental dependence of the absolute UV magnitude, the Ly$\alpha$ luminosity or the rest-frame Ly$\alpha$ equivalent width and interpret this as due to the extreme youth of the system. Given the weak detection of environmental signatures in the $z=3.78$ system presented here, it is likely that the lower redshift protocluster is more evolved. \pcf1\ also contains a larger number of galaxies available for studying the environmental dependencies (165 LAEs in the vicinity of the $z\approx3.78$ cluster versus 28 LAEs in and around the $z\approx6$ cluster), which enables the detection of weaker trends. Also, since the current study investigates a much wider field, it is sensitive to a wider range in environmental densities. It will be of great interest to identify more protocluster regions at a range of redshifts to determine exactly when, how and why environmental dependencies are established. 

\section{Conclusions}

We have obtained deep narrow- and broad-band imaging of a wider field surrounding the $z\approx3.78$ protocluster candidate \pcf1\ discovered by \citet{lee2014}. Combining the new data with the old, we have photometrically identified a total of 165 candidate Ly$\alpha$ emitting galaxies in the region. The large scale structure traced by the LAE candidates shows a very significant overdensity and $2-3$ smaller overdensities and voids. The peak overdensities lie along an elongated, possibly filamentary region of enhanced density stretching more than 170~Mpc. 


Observations using the DEIMOS spectrograph on the W. M. Keck II telescope have measured the redshifts of 79 LAEs, demonstrated the robustness of the LAE selection and confirmed the existence of coherent large scale structure in this region. 
We have measured the velocity dispersion in two overdensities within the field. Using the 41 spectroscopically confirmed galaxies lying within the most significant overdensity, the core of \pcf1, we find the velocity dispersion to be small, $\approx$350~\kms.  
We speculate that \pcf1\ is not dynamically relaxed but is in the process of forming, although a larger sample of spectroscopic redshifts is needed to elucidate kinematic properties of this protocluster region. Based on the current measurements, we estimate the combined mass in this region to be \gtsima $1\times10^{15}~{\rm M_\odot}$, suggesting that the region may evolve to be a Coma-like cluster in the present-day universe.  

The narrow-band \wrc4\ magnitude (which is a proxy for the Ly$\alpha$ flux) shows marginal evidence for a weak environmental dependence: LAE candidates in lower density regions and ones that lie further away from the peak of the central overdensity are fainter (i.e., show weaker Ly$\alpha$ emission). We find that the Ly$\alpha$ luminosity is enhanced by an average factor of 1.35 within the protocluster core. The $WRC4-R$ color (a proxy for the Ly$\alpha$ equivalent width) does not show any trend with environmental density, suggesting that the continuum luminosity may also increase towards denser regions. It is possible that this effect is caused by an enhancement of the star-formation rate in denser regions, but the present data are inadequate to unambiguously answer this question. 

Various projects \citep[e.g., HETDEX;][]{hetdex} are attempting to
use LAEs as tracers of large scale structure and cosmology.  Properly
calibrating the environmental trends and their varying dependence
on redshift will be important for the use of LAEs 
as cosmological and matter density probes.  While narrow-band
filters may be effective ways of searching for high-redshift
protoclusters, these searches have unique biases that have to be
taken into account when estimating overdensities and velocity
dispersions.

\begin{deluxetable*}{lccccccl}
\tablecolumns{8}
\tablecaption{Imaging Summary \label{imobs}}
\tablehead{\colhead{\bf Field} & 
	\colhead{\bf RA} & 
	\colhead{\bf DEC} & 
	\multicolumn{4}{c}{$t_{\rm exp}$ (hours)} & 
	\colhead{\bf Comments} \\
	\colhead{} & 
	\colhead{} & 
	\colhead{} & 
	\colhead{\bf WRC4} &
	\colhead{\bf B$_W$} &
	\colhead{\bf R} &
	\colhead{\bf I} &
	\colhead{}
	}
\startdata
North Field & 14:31:36.1 & 32:36:46 &  8.3 & 5.3 & 6.9 & 4.1 & Lee et al. (2014)\\
South Field & 14:31:36.1 & 32:08:46 & 10.2 & 5.0 & 6.2 & 4.3 & This paper\\
\enddata
\end{deluxetable*}

\begin{deluxetable*}{lcccccccc}
\tablecolumns{8}
\tablecaption{Spectroscopy Summary \label{specobs}}
\tablehead{\colhead{\bf Mask Name} & \colhead{\bf RA} & \colhead{\bf DEC} & \colhead{\bf PA (deg)} & \colhead{\bf $t_{\rm Exp}$} (sec) & \colhead{\bf $N_{\rm LAE}$\tablenotemark{a}} & \colhead{\bf $N_{\rm LAE,z}$}\tablenotemark{b} & \colhead{\bf $N_{\rm LBG}$\tablenotemark{c}} 
}
\startdata
DVPC01 & 14:31:45.76 & 32:28:06.0 & -20.0 & 6292 & 12 & 11 & 24 \\
DVPC02 & 14:32:29.41 & 32:45:26.7 & -20.5 & 7200 & 12 & 11 & 25 \\
DVPC03 & 14:30:55.01 & 32:49:00.0 & -85.0 & 6300 & 11 &  7 & 14 \\ 
DVPC04 & 14:30:46.95 & 32:36:16.9 &  38.5 & 3600 & 13 &  9 & 24 \\
DVPC06 & 14:31:37.93 & 32:19:12.0 & -61.4 & 9000 & 14 & 13 & 24 \\
DVPC07 & 14:31:36.44 & 32:15:36.7 &  37.7 & 7200 & 10 & 10 & 22 \\
DVPC08 & 14:31:18.01 & 32:24:58.6 &  50.2 & 2260 &  7 &  4 & 20 \\
DVPC09 & 14:30:49.98 & 31:54:19.0 &  21.8 & 5400 & 11 & 10 & 13 \\
DVPC10 & 14:32:21.61 & 32:41:33.6 &  51.8 & 2400 & 10 &  6 & 18 \\
\enddata
\tablenotetext{a}{Total number of photometrically selected LAE candidates on the mask.}
\tablenotetext{b}{Total number of spectroscopically confirmed LAEs.}
\tablenotetext{c}{Total number of photometrically selected $B_W$-drop LBG candidates on the mask.}
\end{deluxetable*}

\begin{deluxetable*}{lcccccc}
\tabletypesize{\small}
\tablecaption{Photometry of LAE Candidates \label{laephot}}
\tablewidth{0pt}
\tablehead{
\colhead{Name} & \colhead{$\alpha_{\rm J2000}$} & \colhead{$\delta_{\rm J2000}$} & \colhead{WRC4} & \colhead{(WRC4~$-R$)} &  \colhead{($B_W-R$)} & \colhead{($R-I$)}}
\startdata
  LAE2158 & 217.654430 & 32.89343 & 23.44$\pm$ 0.15 &    -2.29$\pm$ 0.22 & $>$ 2.69 &    -0.16$\pm$ 0.62 \\
  LAE2439 & 218.167860 & 32.88762 & 23.65$\pm$ 0.16 &    -2.47$\pm$ 0.28 & $>$ 2.41 &    -0.30$\pm$ 0.91 \\
  LAE2648 & 218.081310 & 32.88373 & 24.49$\pm$ 0.19 &    -1.70$\pm$ 0.22 & $>$ 2.87 &    -0.52$\pm$ 0.72 \\
  LAE2786 & 218.144920 & 32.87901 & 23.58$\pm$ 0.17 &    -1.71$\pm$ 0.20 & $>$ 2.89 &    -0.57$\pm$ 0.73 \\
  LAE3723 & 217.688470 & 32.86251 & 23.56$\pm$ 0.14 &    -1.26$\pm$ 0.12 & $>$ 3.59 &     0.26$\pm$ 0.19 \\
\enddata
\tablenotetext{}{Full Table is available in machine readable form.}
\end{deluxetable*}

\begin{deluxetable*}{lccccccc}
\tabletypesize{\scriptsize}
\tablecaption{Redshifts and Luminosities \label{laespec}}
\tablewidth{0pt}
\tablehead{
\colhead{Name} & \colhead{$\alpha_{\rm J2000}$} & \colhead{$\delta_{\rm J2000}$} & \colhead{Redshift\tablenotemark{a}} & \colhead{$\rm W^0_{\rm\bf Ly\alpha}$\tablenotemark{b}} & \colhead{$\rm\bf F_{\rm\bf Ly\alpha}$} &  \colhead{$\rm L_{\rm\bf Ly\alpha}$} & \colhead{$\rm\bf L_{1700}$} \\
 & (${\bf ^\circ}$) & (${\bf ^\circ}$) &   & ({\bf \AA}) & (${\bf\rm \times10^{-17}\,erg\,s^{-1}\,cm^{-2}}$) & (${\bf\rm \times10^{42}\,erg\,s^{-1}}$) & ${\rm\bf \times 10^{29} erg,s^{-1},Hz^{-1}}$ }
\startdata
 LAE19829 & 217.936330 & 32.52939 & 3.7784 &   149.4$\pm$ 24.7 &    2.02$\pm$ 0.51 &  2.89$\pm$ 0.73 &  0.49$\pm$ 0.15 \\
 LAE17422 & 217.878970 & 32.57814 & 3.7925 &    38.9$\pm$  1.9 &    1.97$\pm$ 0.25 &  2.85$\pm$ 0.36 &  0.71$\pm$ 0.17 \\
 LAE20822 & 217.950640 & 32.50753 & 3.7837 &    51.6$\pm$  4.5 &    4.56$\pm$ 0.63 &  6.56$\pm$ 0.91 &  3.94$\pm$ 0.24 \\
 LAE24669 & 217.943950 & 32.42646 & 3.7861 &   141.4$\pm$ 13.4 &    4.07$\pm$ 0.60 &  5.86$\pm$ 0.87 &  0.75$\pm$ 0.16 \\
 LAE26289 & 217.997460 & 32.39495 & 3.7850 &    49.8$\pm$  3.3 &    2.23$\pm$ 0.28 &  3.21$\pm$ 0.40 &  0.27$\pm$ 0.14 \\
 LAE28232 & 217.981520 & 32.35929 & 3.7900 &    19.8$\pm$  5.9 &    6.12$\pm$ 2.23 &  8.83$\pm$ 3.22 &  1.65$\pm$ 0.20 \\
 LAE28493 & 217.949280 & 32.35492 & 3.7883 &    39.5$\pm$  5.1 &    1.50$\pm$ 0.30 &  2.17$\pm$ 0.43 &  0.48$\pm$ 0.12 \\
 LAE20817 & 217.931100 & 32.50740 & 3.7745 &    45.2$\pm$  1.1 &    9.29$\pm$ 0.58 & 13.29$\pm$ 0.83 &  2.22$\pm$ 0.30 \\
 LAE24356 & 217.904190 & 32.43267 & 3.7864 &   152.1$\pm$  6.3 &   12.63$\pm$ 0.84 & 18.19$\pm$ 1.21 &  2.39$\pm$ 0.27 \\
 LAE24998 & 217.975860 & 32.41984 & 3.7873 &    27.6$\pm$  5.0 &   11.62$\pm$ 2.68 & 16.75$\pm$ 3.86 &  5.46$\pm$ 0.30 \\
  LAE6017 & 218.028230 & 32.81698 & 3.7768 &    32.7$\pm$  3.0 &    2.00$\pm$ 0.31 &  2.86$\pm$ 0.45 &  1.45$\pm$ 0.15 \\
  LAE6756 & 218.101050 & 32.80177 & 3.7858 &    28.5$\pm$  2.3 &    1.35$\pm$ 0.22 &  1.94$\pm$ 0.32 &  0.44$\pm$ 0.15 \\
  LAE8448 & 218.135330 & 32.76630 & 3.7872 &    33.0$\pm$  0.6 &    6.67$\pm$ 0.37 &  9.61$\pm$ 0.53 &  1.78$\pm$ 0.28 \\
  LAE9469 & 218.094440 & 32.74559 & 3.7824 &    87.0$\pm$  5.7 &    3.37$\pm$ 0.42 &  4.85$\pm$ 0.60 &  0.20$\pm$ 0.60 \\
  LAE9613 & 218.095660 & 32.74258 & 3.7838 &    62.9$\pm$  3.7 &    3.60$\pm$ 0.40 &  5.17$\pm$ 0.58 &  0.68$\pm$ 0.20 \\
 LAE10693 & 218.155290 & 32.71979 & 3.7911 &    39.0$\pm$  1.3 &    2.02$\pm$ 0.23 &  2.92$\pm$ 0.33 &  1.34$\pm$ 0.17 \\
 LAE11077 & 218.119210 & 32.71222 & 3.7762 &    45.3$\pm$  4.2 &    1.78$\pm$ 0.30 &  2.55$\pm$ 0.42 &  0.78$\pm$ 0.16 \\
 LAE13049 & 218.183230 & 32.67070 & 3.7827 &  $>$382.1 &    1.99$\pm$ 0.28 &  2.86$\pm$ 0.40 &  0.15$\pm$ 0.77 \\
 LAE14931 & 218.143630 & 32.63079 & 3.7956 &    78.4$\pm$  7.0 &    1.71$\pm$ 0.29 &  2.48$\pm$ 0.43 &  0.77$\pm$ 0.15 \\
  LAE7236 & 217.832310 & 32.79220 & 3.7997 &   204.3$\pm$ 32.9 &    1.73$\pm$ 0.43 &  2.52$\pm$ 0.63 &  0.13$\pm$12.21 \\
  LAE5952 & 217.624170 & 32.81787 & 3.7795 &    36.9$\pm$  1.8 &    2.14$\pm$ 0.28 &  3.07$\pm$ 0.39 &  1.97$\pm$ 0.18 \\
  LAE6051 & 217.600060 & 32.81902 & 3.7765 &   452.9$\pm$105.3 &    4.24$\pm$ 1.41 &  6.07$\pm$ 2.03 &  0.99$\pm$ 0.17 \\
  LAE5586 & 217.631560 & 32.82481 & 3.7896 &   $>$28.0 &    4.23$\pm$ 3.93 &  6.10$\pm$ 5.67 &  1.02$\pm$ 0.23 \\
  LAE6778 & 217.742150 & 32.80115 & 3.7704 & $>$1067.5 &    5.21$\pm$ 0.38 &  7.43$\pm$ 0.55 &  1.07$\pm$ 0.23 \\
  LAE6929 & 217.653760 & 32.79839 & 3.7925 &    29.0$\pm$  1.2 &    1.31$\pm$ 0.20 &  1.90$\pm$ 0.28 &  0.98$\pm$ 0.14 \\
  LAE7450 & 217.718590 & 32.78743 & 3.8042 &   140.1$\pm$  3.0 &    5.16$\pm$ 0.37 &  7.52$\pm$ 0.54 &  0.99$\pm$ 0.22 \\
 LAE15686 & 217.697990 & 32.61484 & 3.7692 &    70.3$\pm$  8.3 &    1.55$\pm$ 0.31 &  2.22$\pm$ 0.44 &  0.58$\pm$ 0.15 \\
 LAE17078 & 217.615920 & 32.58515 & 3.7704 &    59.9$\pm$  2.8 &    2.31$\pm$ 0.27 &  3.30$\pm$ 0.39 &  0.53$\pm$ 0.16 \\
 LAE19467 & 217.653650 & 32.53646 & 3.7705 &  $>$461.3 &    3.23$\pm$ 0.34 &  4.61$\pm$ 0.48 &  0.44$\pm$ 0.17 \\
 LAE20740 & 217.617750 & 32.50856 & 3.7964 &   137.4$\pm$ 11.3 &    2.20$\pm$ 0.34 &  3.19$\pm$ 0.50 &  0.15$\pm$ 0.15 \\
 LAE27108 & 217.796230 & 32.37852 & 3.7770 &    76.1$\pm$ 13.7 &    1.61$\pm$ 0.43 &  2.30$\pm$ 0.61 &  0.32$\pm$ 0.12 \\
 LAE29640 & 217.772490 & 32.33416 & 3.7650 &  $>$361.0 &    1.15$\pm$ 0.17 &  1.64$\pm$ 0.25 &  0.12$\pm$ 0.09 \\
 LAE30078 & 217.929120 & 32.32691 & 3.7750 &    30.0$\pm$  2.6 &    1.86$\pm$ 0.28 &  2.65$\pm$ 0.40 &  1.08$\pm$ 0.15 \\
 LAE30119 & 217.881820 & 32.32485 & 3.7710 &    49.7$\pm$  1.6 &    4.84$\pm$ 0.34 &  6.91$\pm$ 0.48 &  1.38$\pm$ 0.21 \\
 LAE30737 & 217.945020 & 32.31530 & 3.7773 &    50.1$\pm$  2.1 &    4.15$\pm$ 0.33 &  5.94$\pm$ 0.48 &  0.86$\pm$ 0.20 \\
 LAE30912 & 217.907460 & 32.31213 & 3.7718 &    47.1$\pm$  2.2 &    4.26$\pm$ 0.38 &  6.08$\pm$ 0.55 &  1.00$\pm$ 0.23 \\
 LAE31237 & 217.888110 & 32.30609 & 3.7741 &  $>$694.7 &    8.03$\pm$ 0.46 & 11.48$\pm$ 0.66 &  0.57$\pm$ 0.24 \\
 LAE32172 & 217.989650 & 32.28942 & 3.7815 &  $>$462.7 &    2.26$\pm$ 0.26 &  3.25$\pm$ 0.37 &  0.16$\pm$14.69 \\
 LAE33005 & 218.002150 & 32.27370 & 3.7915 &    90.0$\pm$  5.7 &    2.72$\pm$ 0.33 &  3.93$\pm$ 0.47 &  0.31$\pm$ 0.19 \\
 LAE34929 & 217.974710 & 32.23817 & 3.7806 &  $>$165.3 &    1.34$\pm$ 0.30 &  1.93$\pm$ 0.43 &  0.44$\pm$ 0.14 \\
 LAE27770 & 217.929230 & 32.36759 & 3.7850 &    47.1$\pm$  1.8 &    3.94$\pm$ 0.33 &  5.67$\pm$ 0.47 &  1.53$\pm$ 0.17 \\
 LAE28203 & 217.957140 & 32.36008 & 3.7860 &    58.7$\pm$  4.3 &    1.27$\pm$ 0.20 &  1.83$\pm$ 0.29 &  0.37$\pm$ 0.11 \\
 LAE28570 & 217.959730 & 32.35341 & 3.7855 &    77.1$\pm$  7.4 &    1.86$\pm$ 0.30 &  2.67$\pm$ 0.43 &  0.15$\pm$ 0.12 \\
 LAE32441 & 217.895620 & 32.28475 & 3.7755 &    80.1$\pm$  5.9 &    2.76$\pm$ 0.36 &  3.94$\pm$ 0.52 &  1.19$\pm$ 0.19 \\
 LAE36170 & 217.855830 & 32.21347 & 3.7794 &    79.8$\pm$  6.7 &    2.21$\pm$ 0.33 &  3.16$\pm$ 0.47 &  0.37$\pm$ 0.17 \\
 LAE36822 & 217.837540 & 32.20162 & 3.7794 &  $>$236.2 &    1.12$\pm$ 0.20 &  1.61$\pm$ 0.29 &  0.15$\pm$ 0.11 \\
 LAE37866 & 217.862000 & 32.18189 & 3.8065 &    92.2$\pm$  1.9 &    2.33$\pm$ 0.25 &  3.40$\pm$ 0.37 &  0.55$\pm$ 0.18 \\
 LAE38202 & 217.814790 & 32.17624 & 3.7742 &  $>$345.8 &    1.42$\pm$ 0.21 &  2.04$\pm$ 0.30 &  0.15$\pm$ 0.12 \\
 LAE34275 & 217.867360 & 32.25039 & 3.7780 &   267.8$\pm$ 73.4 &    1.76$\pm$ 0.70 &  2.53$\pm$ 1.01 &  0.15$\pm$ 0.95 \\
 LAE34611 & 217.902750 & 32.24429 & 3.8053 &   105.7$\pm$  3.1 &    6.15$\pm$ 0.45 &  8.97$\pm$ 0.65 &  2.34$\pm$ 0.28 \\
 LAE20938 & 217.940280 & 32.50491 & 3.7812 &  $>$313.1 &    3.87$\pm$ 0.47 &  5.56$\pm$ 0.68 &  0.88$\pm$ 0.19 \\
 LAE21061 & 217.900570 & 32.50249 & 3.7854 &    59.2$\pm$  4.4 &    2.96$\pm$ 0.41 &  4.26$\pm$ 0.59 &  1.09$\pm$ 0.21 \\
 LAE23643 & 217.810770 & 32.44774 & 3.7895 &    30.2$\pm$  3.2 &    1.81$\pm$ 0.32 &  2.61$\pm$ 0.46 &  0.26$\pm$ 0.16 \\
 LAE24206 & 217.817430 & 32.43593 & 3.7796 &  $>$265.3 &    2.25$\pm$ 0.35 &  3.23$\pm$ 0.50 &  0.57$\pm$ 0.14 \\
 LAE26623 & 217.690330 & 32.38755 & 3.8018 &   240.2$\pm$ 24.9 &    2.08$\pm$ 0.36 &  3.02$\pm$ 0.53 &  0.99$\pm$ 0.13 \\
 LAE47338 & 217.759730 & 31.99508 & 3.7625 &    35.7$\pm$  3.0 &    1.94$\pm$ 0.29 &  2.75$\pm$ 0.41 &  0.94$\pm$ 0.20 \\
 LAE48282 & 217.697630 & 31.97667 & 3.7789 &  $>$217.1 &    2.34$\pm$ 0.39 &  3.36$\pm$ 0.56 &  0.18$\pm$ 0.19 \\
 LAE48597 & 217.710130 & 31.97050 & 3.7862 &   266.4$\pm$ 59.3 &    5.24$\pm$ 1.65 &  7.55$\pm$ 2.38 &  0.59$\pm$ 0.24 \\
 LAE49611 & 217.751050 & 31.95267 & 3.7705 &   101.6$\pm$  5.2 &    3.57$\pm$ 0.35 &  5.09$\pm$ 0.50 &  0.43$\pm$ 0.19 \\
 LAE51100 & 217.697250 & 31.92169 & 3.7933 &    79.6$\pm$  3.8 &    3.41$\pm$ 0.33 &  4.93$\pm$ 0.48 &  0.68$\pm$ 0.20 \\
 LAE51157 & 217.694340 & 31.92069 & 3.7965 &   352.3$\pm$109.8 &    2.53$\pm$ 1.13 &  3.67$\pm$ 1.63 &  0.19$\pm$ 0.92 \\
 LAE51900 & 217.718140 & 31.90618 & 3.7966 &   215.7$\pm$ 52.0 &    3.49$\pm$ 1.20 &  5.07$\pm$ 1.74 &  2.31$\pm$ 0.22 \\
 LAE54720 & 217.658190 & 31.85201 & 3.7944 &    38.6$\pm$  3.6 &    1.20$\pm$ 0.21 &  1.74$\pm$ 0.30 &  0.42$\pm$ 0.14 \\
 LAE55496 & 217.644270 & 31.83689 & 3.7822 &    86.0$\pm$  3.4 &    6.60$\pm$ 0.47 &  9.49$\pm$ 0.68 &  0.30$\pm$ 0.27 \\
 LAE56448 & 217.690290 & 31.81790 & 3.7932 &   109.5$\pm$  1.8 &   16.69$\pm$ 0.56 & 24.16$\pm$ 0.81 &  2.67$\pm$ 0.37 \\
  LAE7688 & 218.145700 & 32.78259 & 3.7697 &  $>$103.3 &    2.46$\pm$ 0.72 &  3.50$\pm$ 1.02 &  0.17$\pm$15.17 \\
  LAE9618 & 218.088680 & 32.74251 & 3.7801 &    32.0$\pm$  3.5 &    1.81$\pm$ 0.33 &  2.59$\pm$ 0.47 &  2.09$\pm$ 0.16 \\
 LAE13327 & 218.041390 & 32.66468 & 3.7752 &    95.5$\pm$ 12.4 &    3.18$\pm$ 0.63 &  4.54$\pm$ 0.90 &  0.17$\pm$ 0.22 \\
 LAE13367 & 218.074140 & 32.66393 & 3.8088 &    35.8$\pm$ 11.3 &    1.39$\pm$ 0.59 &  2.03$\pm$ 0.86 &  0.15$\pm$ 0.13 \\
  LAE7621 & 218.207600 & 32.78392 & 3.7640 &  $>$251.5 &    2.23$\pm$ 0.35 &  3.17$\pm$ 0.50 &  0.15$\pm$ 0.22 \\
  LAE9866 & 218.073810 & 32.73668 & 3.7722 &  $>$322.8 &   15.07$\pm$ 1.52 & 21.53$\pm$ 2.17 &  7.66$\pm$ 0.36 \\
 LAE10349 & 218.098620 & 32.72760 & 3.7775 &  $>$184.1 &    1.74$\pm$ 0.35 &  2.49$\pm$ 0.50 &  0.14$\pm$ 0.16 \\
\enddata
\tablenotetext{a}{All redshift measurements are based on gaussian fits to the Ly$\alpha$ emission line.}
\tablenotetext{b}{Rest-frame equivalent width. In the cases where no significant continuum flux is detected, $3\sigma$ lower limits to the rest-frame equivalent width are estimated using 3$\sigma$ upper limits to the continuum flux.}
\end{deluxetable*}

\begin{deluxetable*}{lccccc}
\tablewidth{0pt}
\tablecolumns{6}
\tablecaption{Line-of-Sight Velocity Dispersion Estimates (\kms)\label{veldisp}\tablenotemark{a}}
\tablehead{
  \colhead{Sub-region} & \colhead{Galaxies\tablenotemark{b}} & \colhead{Std.~Dev.} & \colhead{M.A.D.} & \colhead{Gapper} & \colhead{Biweight}
 } 
\startdata
     Core & 41 & 341 $\pm$  30 & 354 $\pm$  73 & 348 $\pm$  29 & 354 $\pm$  35\\
Inner Core& 34 & 350 $\pm$  34 & 250 $\pm$ 101 & 351 $\pm$  39 & 372 $\pm$  60\\
 NE Clump & 16 & 413 $\pm$  58 & 479 $\pm$ 135 & 430 $\pm$  63 & 412 $\pm$  66\\
SW Region &  9 & 573 $\pm$ 125 & 535 $\pm$ 271 & 583 $\pm$ 128 & 557 $\pm$ 205\\
\enddata
\tablenotetext{a}{All velocity dispersions are computed for galaxies in the redshift range $3.77\le z\le3.80$. Uncertainties are estimated based on 1000 bootstrap samples.}
\tablenotetext{b}{Number of galaxies in the sample.}
\end{deluxetable*}

\acknowledgments

We thank the referee for a careful reading of the manuscript and for suggestions that helped improve this paper. 
This paper presents data obtained at the W. M. Keck Observatory (NASA proposal ID numbers 2014A-N116D and 2015A-N142D) and the Mayall 4m telescope of the Kitt Peak National Observatory (NOAO proposal ID numbers 2012A-0454, 2014A-0164 and 2014B-0626). We are grateful to the NASA Keck and NOAO Time Allocation Committees for granting us telescope time and to the staff of the W. M. Keck Observatory and the Kitt Peak National Observatory. We particularly want to thank Greg Wirth, Marc Kassis and Gary Punawai at Keck, and Malanka Riabokin and Doug Williams at the Mayall telescope for their indispensible help with the observations. We also thank Y.-K. Chiang and R. Overzier for useful comments and for sharing the results of their protocluster studies in advance of publication. This work was supported by a NASA Keck PI Data Award, administered by the NASA Exoplanet Science Institute. Data presented herein were obtained at the W.~M.~Keck Observatory using telescope time allocated to the National Aeronautics and Space Administration through the agency's scientific partnership with the California Institute of Technology and the Univeristy of California. The Observatory was made possible by the generous financial support of the W. M. Keck Foundation.  The authors wish to recognize and acknowledge the very significant cultural role and reverence that the summit of Mauna Kea has always had within the indigenous Hawaiian community. We are most privileged to be able to conduct observations from this mountain. We thank NASA for support, through grants NASA/JPL\# 1497290 and 1520350.  AD thanks the Radcliffe Institute for support during the early stages of this project and the Aspen Center for Physics for its hospitality during the month when this paper was completed.  AD's research was supported in part by the National Optical Astronomy Observatory (NOAO) and by the Radcliffe Institute for Advanced Study and the Institute for Theory and Computation at Harvard University. NOAO is operated by the Association of Universities for Research in Astronomy (AURA), Inc. under a cooperative agreement with the National Science Foundation. 

\appendix

\section{Velocity Dispersion Bias in Narrow-Band Surveys}

Narrow-band surveys for Ly$\alpha$ emitters are a very effective method for identifying protoclusters. However, by design, such surveys find only a biased subset of the galaxies in these regions, since such searches select only galaxies that show significant Ly$\alpha$ line emission which lie within the redshift range defined by the wavelength coverage of the narrow-band filter. Spectroscopic follow-up of these protocluster candidate regions using only (or dominantly) the LAE candidates can therefore undersample the full velocity field of the overdensities. While this bias may be an obvious one, it is not often taken into account when velocity dispersions are estimated for protocluster regions using samples dominated by LAE candidates, and hence we describe its effect in this Appendix. 

\begin{figure*}[hbt]
\epsscale{1.0}
\plottwo{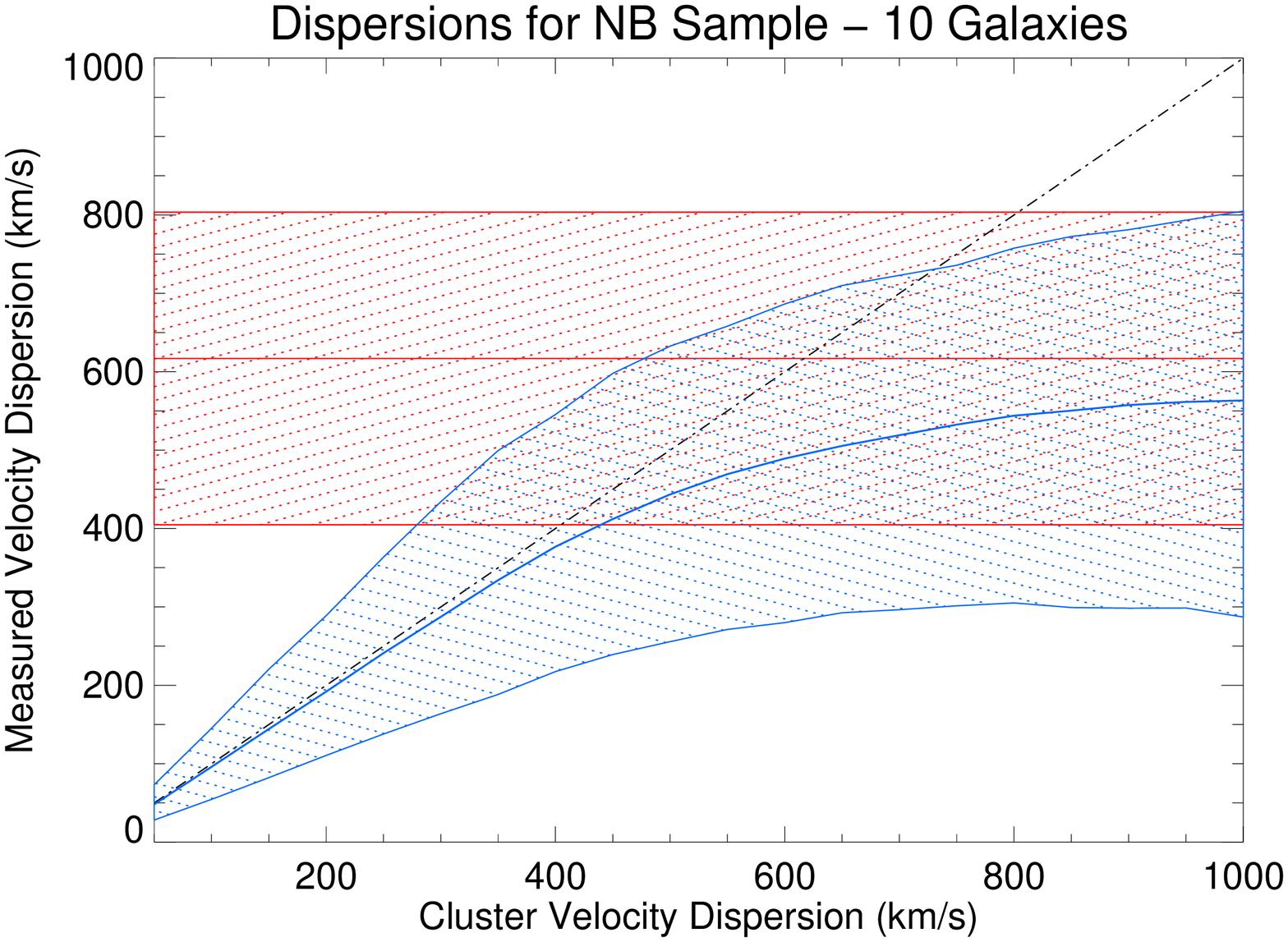}{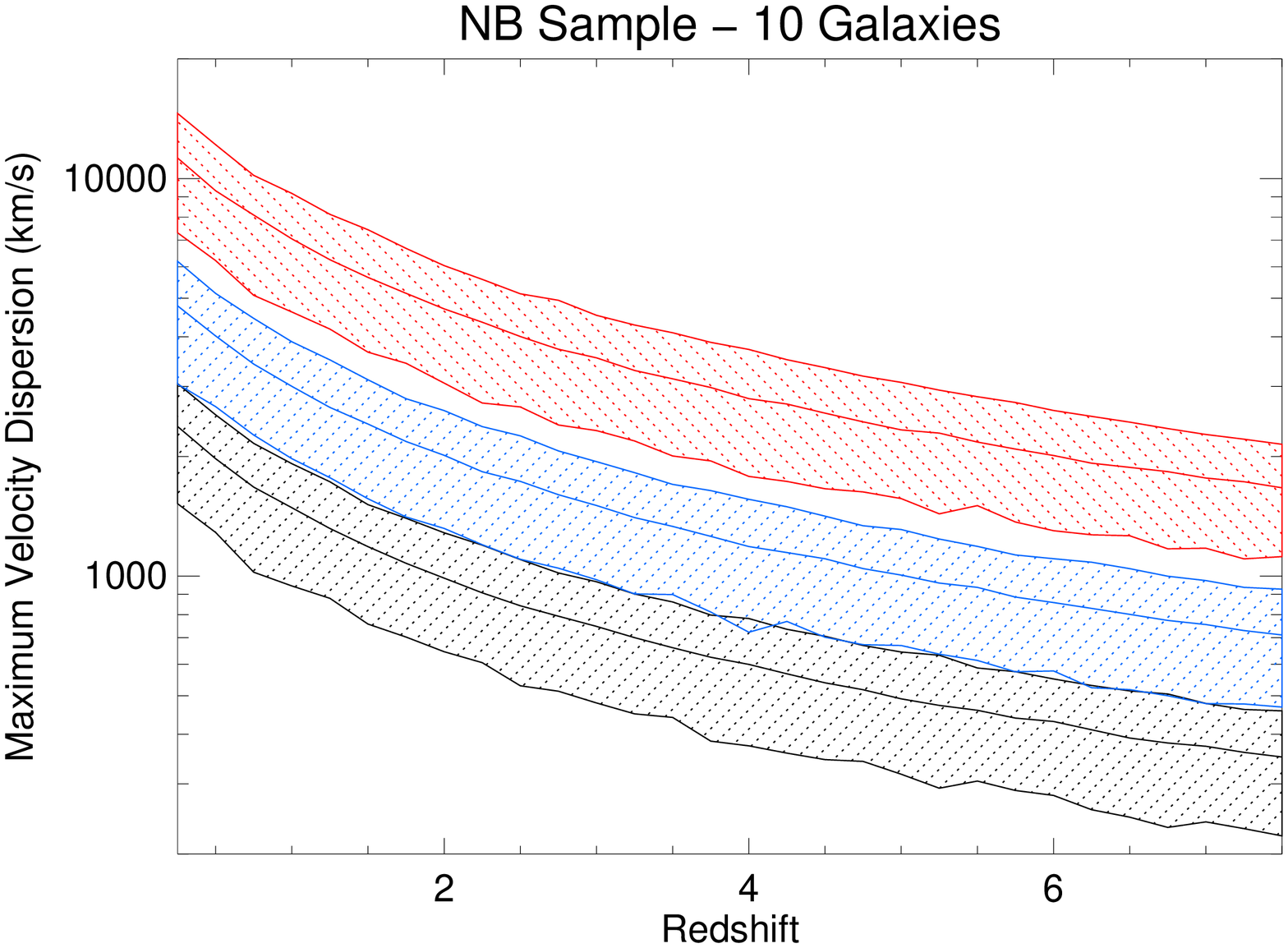}
\plottwo{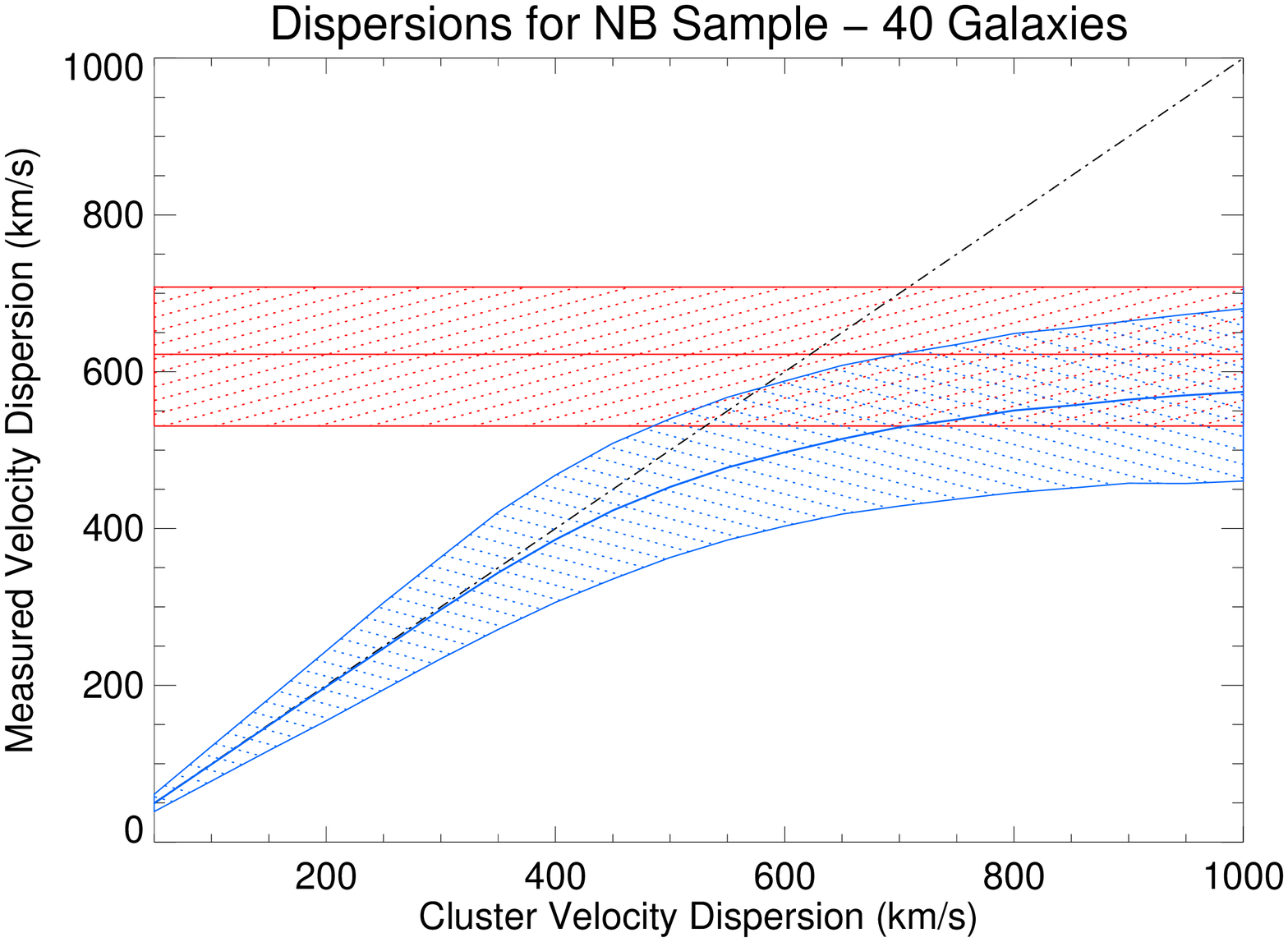}{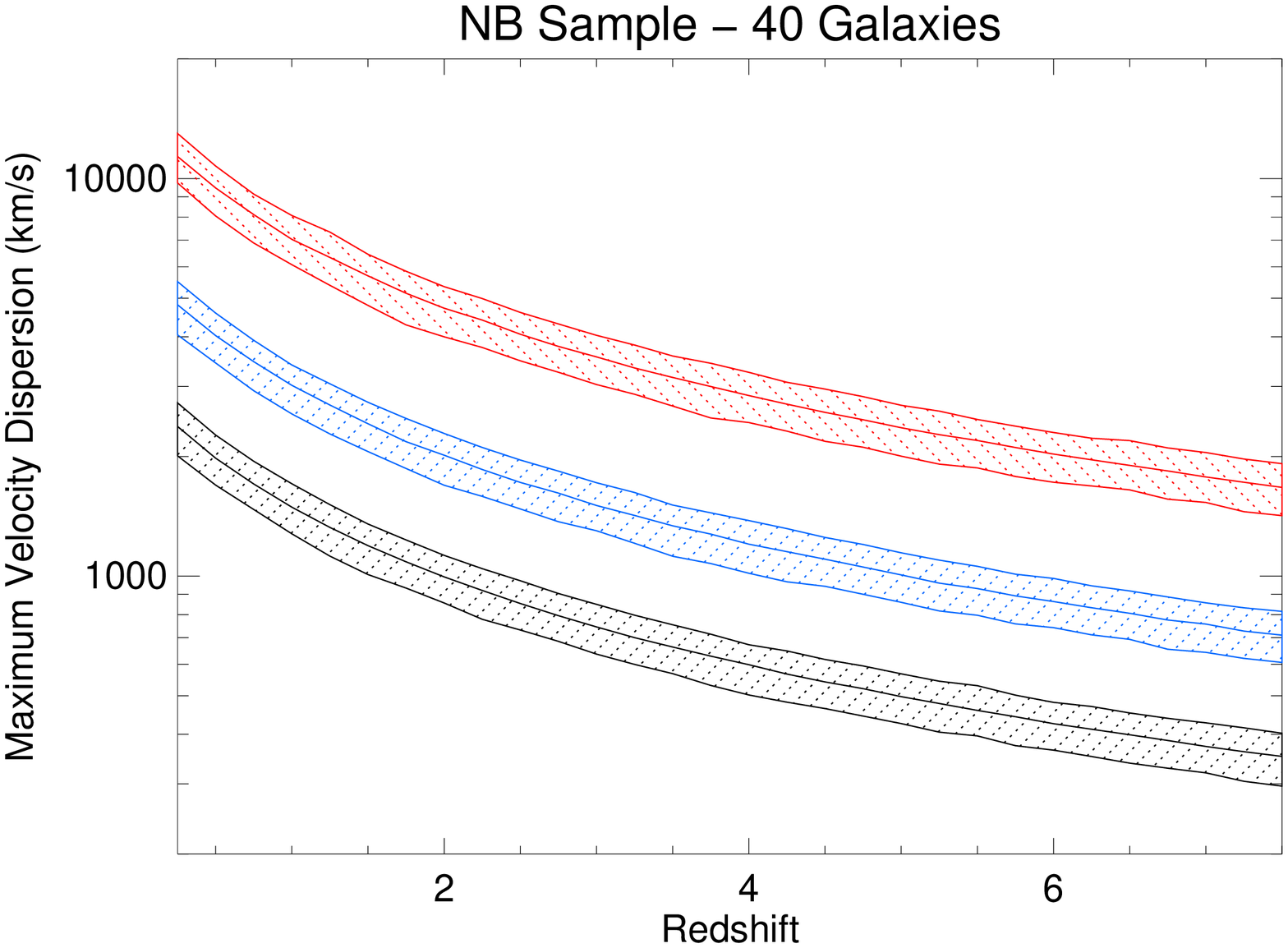}
\caption{The effects of selection biases in narrow-band Ly$\alpha$ emitter samples on the measured cluster velocity dispersions. {\it Left:} The left panels show the bias in the measured cluster velocity dispersion due to the finite bandpass of the narrow band filter. The top and bottom panels show the results for two samples of 10 and 40 LAEs respectively. The solid blue line shows the 
measured velocity dispersion as a function of the true cluster velocity dispersion for a 
sample of emission line galaxy candidates selected using a narrow-band filter of full width 42\AA. 
The cluster velocity profile is assumed to be a gaussian along the line of sight and the narrow-band filter is centered on the cluster redshift and has a tophat transmission profile as a function of wavelength. The diagonal dashed line shows the one-to-one line, where the measured velocity dispersion is identical to the true cluster velocity dispersion. The blue hashed region surrounding the solid black line shows the 95\% scatter range of the measurement. The horizontal solid red line and hashed region represent the maximum velocity dispersion that can be measured if the LAE sample is distributed randomly in redshift along the line of sight and no protocluster was present. 
{\it Right:} The maximum measurable velocity dispersion as a function of redshift 
for emission line galaxy candidates selected using narrow-band filters of fixed widths. 
The black, blue and red solid lines (and hashed regions) show the maximum velocity dispersions (and 95\% confidence limits) for samples selected using filters of width 42\AA, 85\AA, and 200\AA\ respectively. 
\label{sigvsz}}
\end{figure*}

The left panels of Figure~\ref{sigvsz} shows the effect of this bias on samples of 10 and 40 galaxies drawn from a narrow-band survey using a 42\AA\ wide filter. The simulation assumes the most advantageous case, where the  narrow-band filter bandpass is centered on the cluster redshift. The blue lines and hashed region represent the typical value and 95\% scatter range of the measured velocity dispersion for a given true line of sight cluster velocity dispersion. The measured velocity dispersion is a fair representation of the cluster velocity dispersion at dispersion \ltsima~400~\kms, but underestimate the dispersion for higher values of the dispersion. 

The bias is the result of the sample being limited in velocity width due to the narrow-band selection, which effectively results in a maximum possible measureable velocity dispersion for a filter of given width.  Even if the galaxies were distributed randomly along the line of sight (i.e., with no protocluster present), spectroscopy of the 10 galaxy sample could yield (in 95\% of cases) a velocity dispersion of between 400~\kms\ and 800~\kms. In other words, there is a maximum possible measureable velocity dispersion irrespective of whether there is a structure within the field of view, and sampling statistics could result in values as low as 400~\kms\ even for random samples. For a larger sample of 40 galaxies (similar to the sample size in the Core region of the cluster presented in this paper), the uncertainty in the measured velocity dispersion for a random redshift distribution is smaller, between 520 and 700~\kms. 

The right panels shows the maximum measurable velocity dispersion as a function of redshift for three different narrow-band filter widths (42\AA, 85\AA, and 200\AA). These widths are chosen to be representative of filters typically used for Ly$\alpha$ surveys. The maximum measureable velocity dispersion decreases with increasing redshift, as expected for a fixed width filter survey. Since many Ly$\alpha$ surveys are conducted with filters of the same width but different central wavelengths, we encourage caution in deriving any evolution in the velocity dispersions of clusters identified in these surveys when the only members used for the dispersion measurements are the narrow-band selected targets. 

\end{document}